# Statistical modeling of ground motion relations for seismic hazard analysis

Mathias Raschke, Freelancer, Stolze-Schrey Str.1, 65195 Wiesbaden, Germany

*mathiasraschke@t-online.de*

We introduce a new approach for ground motion relations (GMR) in the probabilistic seismic hazard analysis (PSHA), being influenced by the extreme value theory of mathematical statistics. Therein, we understand a GMR as a random function. We derive mathematically the principle of area-equivalence; wherein two alternative GMRs have an equivalent influence on the hazard if these GMRs have equivalent area functions. This includes local biases. An interpretation of the difference between these GMRs (an actual and a modeled one) as a random component leads to a general overestimation of residual variance and hazard. Beside this, we discuss important aspects of classical approaches and discover discrepancies with the state of the art of stochastics and statistics (model selection and significance, test of distribution assumptions, extreme value statistics). We criticize especially the assumption of logarithmic normally distributed residuals of maxima like the peak ground acceleration (PGA). The natural distribution of its individual random component (equivalent to $exp(\varepsilon_0)$ of Joyner and Boore 1993) is the generalized extreme value. We show by numerical researches that the actual distribution can be hidden and a wrong distribution assumption can influence the PSHA negatively as the negligence of area equivalence does. Finally, we suggest an estimation concept for GMRs of PSHA with a regression-free variance estimation of the individual random component. We demonstrate the advantages of event-specific GMRs by analyzing data sets from the PEER strong motion database and estimate event-specific GMRs. Therein, the majority of the best models base on an anisotropic point source approach. The residual variance of logarithmized PGA is significantly smaller than in previous models. We validate the estimations for the event with the largest sample by empirical area functions, which indicate the appropriate modeling of the GMR by an anisotropic point source model. The constructed distances like the Joyner-Boore distance do not work well for event-specific GMRs. We discover also a strong relation between magnitude and the squared expectation of the PGAs being integrated in the geo-space for the event-specific GMRs. One of our secondary contributions is the simple modeling of anisotropy for a point source model.

*ground motion relation, probabilistic seismic hazard analysis, area-equivalence, regression analysis, extreme value statistics, model selection, statistical test, random function*

## 1 Introduction

The level of local seismic impact is estimated for modern building codes and the earthquake resistant design of industrial facilities by probabilistic seismic hazard analysis (PSHA) as a part of seismology and earthquake engineering. Therein, the average annual exceedance frequency of local earthquake ground motion intensity is estimated. An important element of PSHA is the ground motion relation (GMR), which describes the relation between the local ground motion intensity and different event parameters such as the magnitude (Bommer and Abrahamson 2006). It is also called ground motion prediction equation. We prefer the term GMR of Atkinson (2006) because we research an appropriate relation for the PSHA that needs not to be the best prediction and its residual variance for a single event. The previous GMRs are mostly modeled by a statistical regression analysis (Strasser et al. 2009) wherein the event parameters are predictors. Douglas (2001, 2002) provides a good overview of GMRs being published before 2002 and Douglas (2003) gives an excellent overview of all aspects of GMRs like estimation methods or source models. Therein, the physical unit of local ground motion intensity is the peak ground acceleration (PGA) or the maximum of another type of local time history. The parameters of current GMRs are fixed, not event-specific, including the depth parameter. The conditional probability distribution of the local ground motion intensity is generally modeled by the logarithmic normal (log-normal) distribution in the GMR, which implies a normal distribution for the logarithmized ground motion intensity (Joyner and Boore 1993, Strasser et al. 2009). This approach results in unrealistically high estimations of ground motion intensities for low exceedance frequency (Stepp et al. 2001, Abrahamson et al. 2002, Bommer and Abrahamson 2006), which has not been improved by the next generation of GMR (NGA, Abrahamson et al. 2008). Beside this, truncation of the log-normal distribution was suggested to avoid overestimations, but choosing the truncation point is difficult according to Strasser et al. (2008). Therein, statistical estimation methods for truncation points (Raschke 2011) have not been considered. We generally note a lack of consideration of current knowledge of stochastics and statistics in the research of GMR. For example, it is known for a long time that the statistical significance of regression models of GMR should be validated (Joyner and Boore 1981), but many NGAs are not validated in this sense (s. Tab. 2). Beside this, at least the individual random component ($\varepsilon_0$ of Joyner and Boore 1993) of the PGA should follow an extreme value distribution according to the extreme value statistics (Leadbetter at al 1983, Coles 2001). Dupuis and Flemming (2006) have introduced the concept of extreme value statistics into GMR but their paper was not considered any further. In the following section on regression models for GMRs, we criticize important statistical aspects of previous GMR and briefly call arguments for the extreme value distribution of the individual random component in Sec. 3. However, our break with the traditional approaches to GMRs is deeper; in Sec. 4 we mathematically derive the area-equivalence for GMRs in PSHA inspired by equivalences in max-stable random fields (Schlather 2002, Kabluchko et al. 2009). Therein, GMRs are random functions, which include event-specific GMRs and distinction between GMRs for an actual prediction and GMRs for the PSHA. We also introduce an approach to an anisotropic point source model in this section.

In Sec. 5, we numerically research the detectability of the distribution model and the influence of this and other items like the area-equivalence on PSHA. Then, we suggest an estimation concept for our approach to GMR in Sec. 6, including a regression-free estimation of the variance of the individual random component. We partly apply this concept to nine suitable data sets and research the link between the event-specific GMR and the magnitude. Finally, we conclude our results in the last section. We follow here the rules of statistics, use its terms (s. Upton and Cook 2008) and refer to the appendix by the letter A.

## 2 Regression model for GMR

### 2.1 Basic formulation

The GMR is usually formulated by a regression model with the basic formulation (Lindsey 1996; Rawlings et al. 1998; Montgomery et al. 2006)

$$Y = g(\mathbf{X}) + \varepsilon^*, \quad E(Y) = g(\mathbf{X}), E(\varepsilon^*) = 0, V(Y) = V(\varepsilon^*), \tag{1}$$

$Y$ is the predicted variable (response variable, dependent variable, conditional variable or regressand). The regression function $g(\mathbf{X})$ includes a parameter vector $\boldsymbol{\theta}$, which is estimated. The predictors (independent variables, predicting variables or regressors) are the elements of the random vector $\mathbf{X} = (X_1, X_2, \ldots, X_m)$. $E(.)$ are the expectations and $V(.)$ are the variances. The random variable $\varepsilon^*$ is the random component (residual, random term or measurement error) and determines the cumulative distribution function (CDF) $F_y$ of $Y$ under condition of $\mathbf{X}$. If $g(\mathbf{X}) \geq 0$ and $V(\varepsilon^*)$ is proportional to $g^2(\mathbf{X})$, then we write the equivalent formulation

$$Y = \varepsilon\, g(\mathbf{X}), \quad \varepsilon = 1 + \varepsilon^*, E(Y) = g(\mathbf{X}), E(\varepsilon) = 1, V(Y) = V(\varepsilon)g^2(\mathbf{X}). \tag{2}$$

We prefer this formulation for GMRs, wherein $Y$ is the PGA or something similar, because the expectation is a very important characterization of a random variable and $\varepsilon$ can be neglected under certain conditions (Sec. 4.1). If $Y \geq 0$, then we can logarithm and formulate the popular model for GMRs (Douglas 2001, Abrahamson et al. 2008)

$$\ln(Y) = g^*(\mathbf{X}) + \xi, E(\ln(Y)) = g^*(\mathbf{X}), E(\xi) = 0, V(\ln(Y)) = V(\xi). \tag{3}$$

It is assumed for most GMRs for PSHA that $\xi$ is normally distributed (Joyner and Boore 1993, Strasser et al. 2009). This implies a model according to Eq.(2) with log-normally distributed $\varepsilon$. The link between Eq.(2,3) is (Johnson et al. 1994, Eq.( 14.8))

$$E(Y) = g(\mathbf{X}) = \exp(g^*(\mathbf{X}) + V(\xi)/2) \text{ and} \tag{4a}$$

$$V(Y) = \exp(2g^*(\mathbf{X}))\exp(V(\xi))(\exp(V(\xi)) - 1), \tag{4b}$$

$$\varepsilon = \exp(\xi) - \exp(V(\xi)/2). \tag{4c}$$

We apply Eq.(3,4) simultaneously even if $\varepsilon$ is not exactly log-normally distributed (Johnson et al. 1994, Eq.(12.67) with $\xi_{Johnson} \approx 0$). A typical formulation for a GMR is (Douglas 2002)

$$g^*(\mathbf{X}) = \theta_0 + \theta_1 m - \theta_2 r - \theta_3 \ln(r) + \theta_s x_s + ..., \quad r = \sqrt{d^2 + h^2}, \theta_1 > 0, \theta_2 \geq 0, \theta_3 \geq 0 \qquad (5)$$

with predicting variables magnitude $m$, source distance $d$ (and $r$) and indicator variable $x_s$ and its parameter $\theta_s$ for the site condition. The source depth is considered by $h$ and can be a parameter or an event-specific predictor. There are many variants and extensions for $g^*(\mathbf{X})$ (Douglas 2002, Abrahamson et al. 2008).

## 2.2 Random components, estimation methods and errors

The random components $\varepsilon$ resp. $\xi$ are independently and identically distributed (iid) variables and the predictors are measured exactly in simple regression models. For such cases, the least squared (LS) estimation can be applied, which is equivalent to the maximum-likelihood (ML) estimation for normally distributed residuals (s. Rawlings et al. 1998, p. 77). This is not popular in seismology, e.g. Castellaro et al. (2006) incorrectly claim that the residuals have to be normally distributed for the LS regression. The LS method has often been used for GMRs and is extended to random components that are not iid. Douglas (2003, Sec.11) gives an overview of approaches from before 2003. The two most important approaches seem to be the one and two stage regression method with the following random components (Joyner and Boore 1993; with assumption of normal distribution)

$$\xi = \xi_E + \xi_S + \xi_0, \quad E(\xi_E) = E(\xi_S) = E(\xi_0) = 0, \qquad (6)$$

wherein $\xi_E$ is event-specific, $\xi_S$ is site-specific and $\xi_0$ has an individual realization for each site (station) and event. We prefer the product formulation according to Eq.(2) with

$$\varepsilon = \varepsilon_E \varepsilon_S \varepsilon_0 \varepsilon_Q, \quad E(\varepsilon) = E(\varepsilon_E) = E(\varepsilon_S) = E(\varepsilon_0) = E(\varepsilon_Q) = 1, \varepsilon_Q = g_{actual}(\mathbf{X}) / g_{equivalent}(\mathbf{X}), \qquad (7)$$

wherein the additional pseudo-random component $\varepsilon_Q$ (resp. $\xi_Q$) results from the ratio between actual and equivalent function $g(\mathbf{X})$ according to Sec.4. A general distribution assumption is not required, but it is obvious that $\varepsilon_Q$ has a finite upper bound and a lower bound larger than 0 for a fixed distance $d$. That is one reason why $\varepsilon_Q$ cannot be log-normally distributed.

In other GMRs, the component $\varepsilon_S$ resp. $\xi_s$ had been replaced by site-specific predictors $x_s$ in Eq.(5). But there is no proof that one additional predictor can completely replace $\varepsilon_S$ resp. $\xi_S$ and we doubt this because site response is very complex. Independent of this, one condition of the regression models of Joyner and Boore (1993) is that predictors $m$ and $r$ do not include a measurement error. However, magnitudes are not measured exactly. Rhoades (1997) considers the known variance of the seismological magnitude estimation in his regression analysis for GMR. It is not considered that this known error needs not to be the only one. The actual error of the seismological estimation can be higher according to Giardini (1984). Arguments for this: The source mechanism influences the ground motion (Campbell 1981 and 1993, Crouse and McGuire 1996, Sadigh et al. 1997). This acts like a measurement error of magnitudes in the GMR. An application of fewer classes of source mechanism would reduce but not eliminate it. Furthermore, the inter-event variability $\xi_E$ can be interpreted as an error in magnitudes, because the seismological magnitudes can be exact for a certain aspect of the rupture process but do not need to be exact for the GMR. The actual magnitude of GMR could be a non-measurable, latent variable, which is estimated by common magnitudes with an error in the sense of statistical error-in models (Cheng and Ness 1999, Sec.1.1). The considerable differences between the estimated residual variances of GMRs for one sample of PGAs but for different magnitude scales (see e.g. Atkinson and Boore 1995, Tab.5) support this assumption. Additionally, the magnitudes of the analyzed sample could be from different scales (e.g. Bommer et al. 2007), which acts like a measurement error.

The source-to-site distance $d$ is also treated as exactly measured predictor. But it should include an error because there are many definitions for this distance (Douglas 2003, Sec.9). How could it be possible that all these measures for the same physical aspect act without a measurement error? Moreover, the distances are determined by the seismological source estimation, which also includes errors. Even if parameters of this error would be known, it would be difficult to consider it in a regression analysis (personal communication with [pcw] Douglas, spring 2013). Beside this, the influence of the source depth is often reduced to a fixed parameter for a defined class of earthquakes (e.g. shallow events). In other GMRs (Ambraseys and Bommer 1991), $h$ is the seismological epicenter depth. But neither is the influence of the source depth the same for every earthquake nor is the seismological depth exactly measured. In both cases, a kind of measurement error is neglected. Furthermore, it is assumed for current GMR that the parameter vector $\boldsymbol{\theta}$ of the GMR is the same for each event (Joyner and Boore 1993, Abrahamson et al. 2008).

There are more estimation methods for a regression model (e.g. Rawlings et al. 1998, Sec.10; Stromeyer et al. 2004). The models for unknown measurement errors of predictors (Cheng and Ness 1999, Sec.4) are not applied for GMR as far as we know. Beside this, the aspect of estimating the estimation errors of the regression parameters is not considered in all approaches. These standard errors can be easily estimated for a simple linear LS regression with iid random components (Rawlings et al. 1988, Sec.4.6). But it is more difficult for models with random effects. Joyner and Boore (1993) applied the Monte Carlo simulation to estimate the estimation error, Rhoades (1997) has computed this standard errors using the likelihood function. Chen and Tsai (2002) also give a method to estimate the standard error. But Abrahamson and Young (1992) do not give any advice for this issue regarding their procedure. We draw attention here to the fact that an estimation error can be computed by the Jackknife technique (Quenouille 1956, Efron 1979). This also applies for clustered data according to Raschke (2012, 2013), as is the case for the mixed effects. The estimation error can be applied directly to construct the confidence range and verify the statistical significance of a predictor and its parameter.

### 2.3 The danger of over-parameterization

We could explain the entire variance of a predicted variable $Y$ or $ln(Y)$ by a regression model if we use a large number of predictors and related parameters, although not all predictors have an actual influence (Rawlings et al. 1993, Fig.8.2). The question is: how can we distinguish between significant and insignificant predictors and/or parameters? Different statistical tools can solve this problem. The first one is the significance test for the regression parameters $\theta_l$ in $g(\mathbf{X})$ resp. $g^*(\mathbf{X}) = \ldots + \theta_i X_i$. We test here if $\theta_i \neq 0$, $\theta_i \leq 0$ or $\theta_i \geq 0$ for a defined significance level $\alpha$ (5% is often used and recommended here). The last two variants are applied when physical reason bounds the influence of a predictor, e.g. a larger magnitude should be related to a larger PGA. In this case, we can be sure with a probability of 100%-$\alpha$ that the actual parameter $\theta_l$ does not have a contrary sign. The smaller $\alpha$ is, the more rigorous is the test. The t-test is such a test (Rawlings et al. 1998, Sec.1.6 and 5.3), which has seldom been applied for GMR, e.g. by Joyner and Boore (1981), Molas and Yamazaki (1995) and Ambraseys et al. (2005, pcw Douglas March 2013). Note that the classical t-test cannot be applied without modification or acceptance of inaccuracies to the case of mixed effects (clustered data). The significance of a published GMR is also examined implicitly by published standard errors of the parameter estimation. If the related quantile, corresponding with $\alpha$, is not smaller/larger than 0, then it implies statistical significance. This is the case for the estimations of Joyner and Boore (1993, Tab.3) and Rhoades (1993, Tab.1). Applying model selection criterions in the model building (Rawlings et al. 1998, Sec.7) is a further possibility for guaranteeing the statistical significance, e.g. the Akaike information criterion (AIC) or the Bayesian information criterion.

The significance has to be verified for each statistical model. Otherwise, danger of over-parameterization arises. This problem applies to a considerable amount of GMRs; we list 15 examples in Tab.1. This is also an issue in other researches (e.g. Raschke and Thürmer 2008).

Tab.1: Examples of GMRs without sufficient validation of significance/model selection ([*]refer to examples)

| # | Reference | Description |
|---|-----------|-------------|
| 1 | Youngs et al. 1995 | The dependency of the residual variance on the magnitude had been tested by a likelihood ratio test but a test for the parameters of the primary model, the GMR, is not mentioned. |
| 2 | Douglas 2002[*], 2003[*] | Many listed GMRs have been developed without a significance test or statistical model selection. |
| 3 | Chen and Tsai 2002 | GMR (Eq.(9), Tab.2) use magnitude-related parameters $\theta_5$ and $\theta_6$, which have a very large standard error; contrary signs are relatively likely. |
| 4 | Boore and Atkinson 2007 | The NGAs have been developed without a significance test (pcw Boore). |
| 5 | Enescu and Enescu 2007 | The anisotropic GMR for Vrancea region (Romania) include 90 parameters (Tab.1) without a significance test or something similar. |
| 6 | Sørensen et al. 2010 | The anistropic, macroseismic GMR for Vrancea region (Romania) include more than 30 parameters without a significance test or something similar (pcw Stromeyer). |
| 7 | Bommer et al. 2007 | No standard errors are given for the ten regression parameters (Tab.2), no significance test or something similar is mentioned. |
| 8 | Campbell and Bozorgnia 2008 | A test is not mentioned for the NGA but applied with $\alpha$=10% (pcw Campbell). The test variant is not very strict; the accepted probability that one parameter is insignificant is $0.65=1-(1-0.1)^{10}$ in case of 10 independent parameters. 16 parameters are estimated (Tab.2). |
| 9 | Chiou and Youngs 2008 | The dependency of the residual variance on the magnitude had been tested by a likelihood ratio test but a test for the parameters of the primary model, the GMR, is not mentioned. |
| 10 | Abrahamson and Silva 2008 | No statistical test is mentioned and the standard errors of the parameter estimation of the NGA with many parameters are not given. A PEER report is referred to for the last one, but this report is not accessible. |
| 11 | Idriss 2007 | No statistical test is mentioned and the standard errors of the parameter estimation of this NGA are not given. |
| 12 | Al Atik et al. 2010[*] | Five references are listed in Tab. 3 and 4 for decomposition of the residual spreading. We examined the references Atkinson (2006; pcw Atkinson), Chen and Tsai (2002), Lin et al. (2011, pcw Lin) and Morikawa et al. (2008). Therein, statistical significance was not validated. |
| 13 | Anderson and Uchiyama 2011 | They have investigated site and path effects in GMR without a validation of significance. The range of only one standard error of site and path related mean residuals in Fig.6 mostly include 0. This is an indication of insignificance. |
| 14 | Scherbaum et al. 2004 | A criterion of model selection for GMRs has been formulated without proof or derivation according to the rules of statistics (s. our Sec. A5). It is applied in other researches, e.g. by Stafford et al. (2008). Therein, the number of parameters is not considered. |
| 15 | Kaklamanos and Baise 2010 | A criterion of model selection for GMRs is introduced without proof or derivation according to the rules of statistics. Therein, the number of parameters is not considered. |

## 2.4 The test of the distribution assumption

Any statistical distribution model should be validated (D'Augustino and Stephens 1986). This also applies to the residual distribution of a GMR in PSHA although a distribution assumption is not necessary for the LS regression. A powerful goodness-of-fit test is the best method of examining the distribution assumption, as the Anderson-Darling (AD) test for a normal distribution (Landry and Lepage 1992). Contrary to the aforementioned t-test, the test is the more rigorous the larger the selected significance $\alpha$ is. There are such tests for different distribution functions with estimated parameters (Stephens 1986). If all parameters are known, then the distribution is fully specified and the classical Kolmogorov-Smirnov (KS) test can be applied. If the KS test is applied to estimated parameters, then the test does not work (Raschke 2009). If there is not an applicable goodness-of-fit test for the distribution type used, then a quantile plot (q-q plot) can be used for a visual, qualitative test as done by Dupuis and Flemming (2006) for residuals with a mixed, non-normal distribution. However, there is no objective criterion for rejecting the distribution hypothesis in this case. A histogram is a kind of parameter-free distribution model; but it is not a tool for validating a distribution model (not mentioned by D'Augustino and Stephens 1986) because there is no objective criterion for rejection and there are many possible histograms for a sample. We state that the assumption of normally distributed $\xi$ resp. its components in Eq.(3) is often not correctly validated for GMRs. For example, Ambraseys and Bommer (1991), Ambraseys and Simpson (1996), Ambraseys et al. (1996), Atkinson and Boore (1995), Spudich et al. (1999), Douglas and Smit (2001), Atkinson (2004) and Kalkan and Gülkan (2004) have neither assumed nor tested a distribution model. Beside this, the assumed normal distribution of $\xi$ has been tested by the inappropriate KS test in other studies (e.g. McGuire 1977, Campbell 1981, Abrahamson 1988, Monguilner et al. 2000, Restrepo-Velez and Bommer 2003). The quantile plot (e.g. Chang et al. 2001, Bommer et al. 2004) and the histogram (e.g. Atkinson 2006, Beyer and Bommer 2007, Morikawa et al. 2008) have also been applied to validate the normal distribution although these methods are not appropriate. Note that even though $\xi$ seems to be normally distributed, it needs not to be (s. Sec.5). The inappropriate test of a distribution assumption is also a problem in other researches, e.g. of flood hazard in Germany (Raschke and Thürmer 2008).

## 3  The distribution of the maximum of a random sequence

The popular assumption for GMR for PGAs, that all random components $\varepsilon_{...}$ are log-normally distributed ($\xi_{...}$ normally distributed, Joyner and Boore 1993, Strasser et al. 2009), is in contradiction to the extreme value theory. According to this special field of stochastics and statistics, the maximum of a sample $Y=Max\{Z_1,Z_2,..,Z_i,...,Z_n\}$ of iid random variables has a generalized extreme value distribution in most cases (GED, Fisher and Tippett 1928, Gnedenko 1943, Beirlant et al. 2004, de Haan and Ferreira 2006). The maximum of a sequence of non-iid random variables also has a GED under some weak conditions (Leadbetter et al. 1983, Part II and III, Falk et al. 2011, Part III). Its CDF is written with the extreme value index $\gamma$ (shape parameter), scale parameter $\sigma$ and location parameter $\mu$

$$G(y) = \exp\left(-\left(1-\gamma(y-\mu)/\sigma\right)^{-1/\gamma}\right) \quad \gamma \neq 0,\ 1-\gamma(y-\mu)/\sigma > 0, \tag{8a}$$

$$G(y) = \exp\left(-\exp\left(-(y-\mu)/\sigma\right)\right), \quad \gamma = 0, \tag{8b}$$

with the Fréchet domain for $\gamma>0$, the Weibull domain for $\gamma<0$ and the Gumbel domain for $\gamma=0$. The latter one is also called the Gumbel distribution. A fundamental property of the GED is max stability. This means that the maximum $max\{Y_1, Y_2\}$ of extreme value distributed variable $Y$ is again extreme value distributed with the same extreme value index $\gamma$. If $Y$ would be log-normally distributed, then $max\{Y_1,Y_2\}$ would not be log-normally distributed. Neither the normal nor the log-normal distributions are max-stable. This problem is typical for the combination of the two horizontal components *(Y₁,Y₂)* of the earthquake record, it could be defined by a maximum *max{Y₁,Y₂}* (Douglas 2003, Sec.6). All combinations with a maximum definition of Douglas (2003, Sec.6, #2, 4, 5) result in a classical extreme value for which the GED is the natural distribution. The log-normal assumption for random component $\varepsilon_0$ is wrong in this case according to the state of the art of stochastics and statistics. We consider here only maxima. In case of other combinations of the horizontal components, it is also unlikely that random component $\varepsilon_0$ becomes log-normally distributed. We have investigated the distribution of the combination arithmetic mean, geometric mean and vectorial addition of Gumbel distributed components $\varepsilon_{01}$ and $\varepsilon_{02}$ numerically (Sec.A2). The log-normal assumptions are rejected.

The argument of missing max-stability of the log-normal assumption also applies for sub-sections of the ground motion time history. If the sub-maxima of not overlapping sub-sections of the time history are log-normally distributed, then the maximum of the entire time history cannot be log-normally distributed (exception: all sub-maxima are identical). Log-normal assumptions would also contradict all our experiences with extreme values (Hüsler et al. 2011, Raschke 2011, 2012, 2013).

We briefly investigate the possible domain of attraction for PGAs and analyze the tail of three acceleration time histories (Fig.1). The tails are exponentially distributed, which indicates the Gumbel domain of attraction for the maxima of the accelerations (s. Coles 2001, Sec.4). Besides, Dupuis and Flemming (2006) have estimated a GMR with GED for the residuals of PGA with extreme value index $\gamma \approx 0$, which also indicates the Gumbel domain.

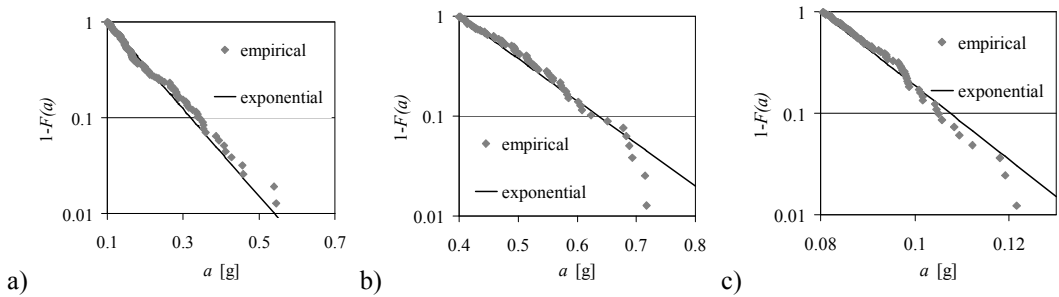

Fig.1: Tails of the time series of ground acceleration *a* of the PEER strong motion database (PEER 2010): a) station: CDMG 24278, component: 090, earthquake: Northridge earthquake 01/17/94, b) station: ARAKYR, component: 090, earthquake: GAZLI 5/17/76, c) station: SMART1 I07, component: NS, earthquake: TAIWAN

# 4   GMR in the PSHA as random function in geo-space

## 4.1   *GMRs as random functions in space*

A random function is a function randomly selected from a set of functions (population). Schlather (2002, Theorem 1 and its proof; we use different symbols) applies a measurable random function $W(\mathbf{s}\text{-}\mathbf{t}) \geq 0$ to construct a max-stable process. This max-stable process $max(Y(\mathbf{s}_i))$ refers to the maxima at site $\mathbf{s}$ from all point events $i$ with local event intensity $Y(\mathbf{s}_i) = m_{oi}W(\mathbf{s}_i\text{-}\mathbf{t})$. Therein, $\mathbf{t}$ is a source allocation resp. the source point in the sense of PSHA (not necessary a point source), being part of a homogeneous Poisson process in the geo-space and at a moment scale $m_o \geq 0$ with density $m_o^{-2}$. Its (annual) maximum $max(Y(\mathbf{s}_i))$ has the CDF $G(y) = \exp(-\lambda(y))$ with annual exceedance function (AEF) $\lambda(y)$ of annual average frequency of exceedance $Y(\mathbf{s}) \geq y$ according to the limit law for extremes (Coles, Sec.7.3). This distribution is equivalent for different sets of random functions if their expectations $E(V_o)$ of the volume

$$V_o = \int_s W(\mathbf{s}-\mathbf{t})ds$$ are equal. This construction can be interpreted as homogeneous seismicity

with $m_o W(\mathbf{s}\text{-}\mathbf{t}) = g(\mathbf{X})$. The spatial distance $\mathbf{s}\text{-}\mathbf{t}$ is part of the distance elements of the predicting vector $\mathbf{X}$ beside the event parameters in sub-vector $\mathbf{X_E}$ of $\mathbf{X}$. It is scaled by the earthquake magnitude with $m = ln(m_o)$ without influence on the shape of $W(\mathbf{s}\text{-}\mathbf{t})$ resp. $g(\mathbf{X})$. The magnitude is exponentially distributed without an upper bound. According to Theorem 1 of Schlather, the AEF $\lambda(y)$ is equal for different GMRs if the expectation $E(V_o)$ is equal for any fixed event parameters like magnitude $m$ and $h$. Furthermore, we can extend this equivalence to $m_o W(\mathbf{s}\text{-}\mathbf{t}) = \varepsilon(\mathbf{s})g(\mathbf{X})$ according to Theorem 2 of Schlather (2002; pcw Kabluchko 2012) if the random component $\varepsilon$ has the expectation $E(\varepsilon) = 1$. This includes that for an event is $\int_s g(\mathbf{X})ds \approx \int_s \varepsilon(\mathbf{s})g(\mathbf{X})ds$ if the variance and the spatial correlation of $\varepsilon(\mathbf{s})$ is relative small. In all cases, we can interpret a GMR as a random function in space being an element of a set (population). Different sets of GMRs act equivalent if the expectations $E(V_o)$ are equivalent. Note that the variance and the distribution type of $\varepsilon$ have no influence on $\lambda(y)$. This could be the reason, why Cornel (1968) did not explicitly consider a random component in his GMR for the PSHA with exponentially distributed magnitudes without upper bound – he did not need and could have find this out by numerical researches. Non-mathematicians can check all results in the same way.

This equivalence of GMRs works only for exponentially distributed magnitudes without upper bound. We introduce the area function $K(y)$ to derive a general equivalence of GMRs being independent of the magnitude distribution.  For this purpose, we consider at first GMR $g(\mathbf{X})$ in a simple one-dimensional geo-space as shown in Fig.2b. We use an example with two maxima of $g(\mathbf{X})$ to demonstrate the general application of this equivalence. For fixed event parameters and a fixed value $y$, the GMR covers a certain area of all points with $y \leq g(\mathbf{X})$. The area function $K(y)$ is for homogeneous site conditions

$$K(y) = \int \mathbf{1}\big(g(\mathbf{X})\big)d\mathbf{t}, \quad \mathbf{1}\big(g(\mathbf{X})\big) = 1 \quad for \quad y \leq g(\mathbf{X}), \quad otherwise \ \mathbf{1}\big(g(\mathbf{X})\big) = 0 \ . \qquad (9)$$

The first derivation is the related area density measuring the amount of points with $y = g(\mathbf{X})$

$$k(y) = -dK(y)/dy \ . \qquad (10)$$

The area function is defined according to Fig.2a and b for a fixed source point $\mathbf{t}$. But we could also fix $\mathbf{s}$ and draw $g(\mathbf{X})$ at $\mathbf{t}$ although it acts at site $\mathbf{s}$. We reflect the GMR in this way for 1D geo-space, as shown in Fig. 2c. We have an equivalent formulation for $K(y)$ with

$$K(y) = \int \mathbf{1}\big(g(\mathbf{X})\big)d\mathbf{s}, \quad \mathbf{1}\big(g(\mathbf{X})\big) = 1 \quad for \quad y \leq g(\mathbf{X}), \quad otherwise \ \mathbf{1}\big(g(\mathbf{X})\big) = 0 \ . \qquad (11)$$

The reflection becomes complicated for the two-dimensional geo-space, but Eq.(9-11) still apply. We can illustrate the reflection for an isoline with fixed $y = g(\mathbf{X})$, as shown in Fig.2d for an anisotropic GMR for a point source and an isotropic GMR for a line source in Fig.2e.

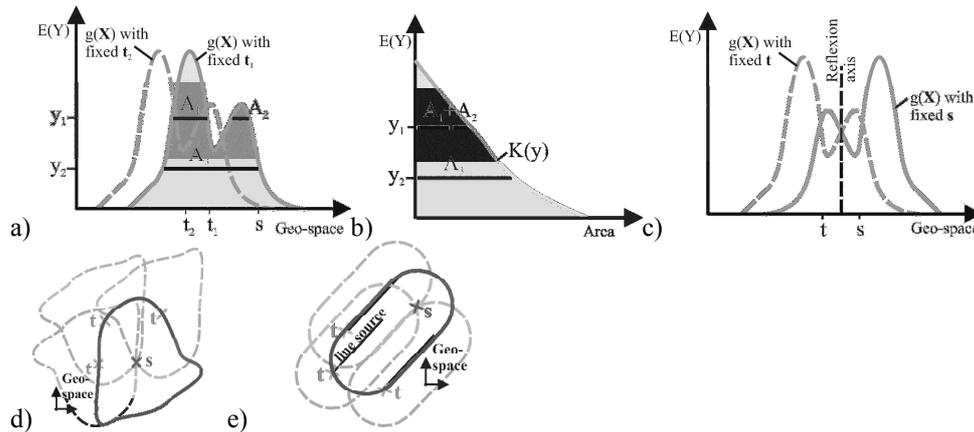

Fig.2: GMR and area-equivalence: a) $g(\mathbf{X})$ in a 1D geo-space, b) resulting area function $K(y)$, c) $g(\mathbf{X})$ for fixed source point $\mathbf{t}$ and reflected version for fixed $\mathbf{s}$, d) 2D geo-space with isolines of $g(\mathbf{X})$ for different $\mathbf{t}$ (light gray) and reflected isoline (dark gray) for fixed $\mathbf{s}$ for anisotropic point source model, e) as $d)$ but for a line source

Now, let us assume the case of homogeneous seismicity: each point $\mathbf{t}$ in the geo-space represents a source allocation with equivalent occurrence intensity $\nu$, equivalent $g(\mathbf{X})$ with $\mathbf{X} = (\mathbf{s}, \mathbf{t}, \mathbf{X}_E)$ with event parameter $\mathbf{X}_E = (m, h, x_l)$ and its multivariate probability density function $f_E$. The distances $d$ resp. $r$ of the GMR are determined by $\mathbf{s}$, $\mathbf{t}$, $h$ and the source model. We formulate for the AEF $\lambda(y)$ of annual average frequency of exceedance $Y(\mathbf{s}) > y$

$$\lambda(y) = \nu \int_{\mathbf{X}_E} \int_{\mathbf{t}} f_E(\mathbf{X}_E)\big(1 - F_y(y; E(Y(\mathbf{s})) = g(\mathbf{X}), V(Y(\mathbf{s})) = g^2(\mathbf{X})V(\varepsilon)\big)d\mathbf{t}d\mathbf{X}_E, \ \mathbf{X} = (\mathbf{X}_E, \mathbf{s}, \mathbf{t}) \ . \qquad (12)$$

Therein, the CDF $F_y$ is parameterized by its expectation $E(Y(s))$ and variance $V(Y(s))$ according to Eq.(2). $V(\varepsilon)$ can be influenced by $\mathbf{X}_E$ but does not include $\varepsilon_Q$. Eq.(12) is oriented on the absolute probability integral of McGuire 1995, but there are many equivalent formulations. One includes a replacement of the integration on **t** by the area density $k(y)$ and the integration on $y=g(\mathbf{X})$ with

$$\lambda(y) = \nu \int_{\mathbf{X}_E} \int_z k(z|\mathbf{X}_E) f_E(\mathbf{X}_E) \left(1 - F_y\left(y; E(Y(\mathbf{s})) = z, V(Y(\mathbf{s})) = z^2 V(\varepsilon)\right)\right) dz d\mathbf{X}_E, \tag{13}$$

because the integration in the geo-space in Eq.(12) is nothing else than a computation of the amount of points with $y=g(\mathbf{X})$ in the sense of measure theory (Billingsley 1995, Chap.2). Now, it is obvious that two GMRs $g_1(\mathbf{X}) \neq g_2(\mathbf{X})$ result in equivalent hazard with $\lambda_1(y)=\lambda_2(y)$ if the area density is equal with $k_1(y)=k_2(y)$ resp. $K_1(y)=K_2(y)$. Note, all other components in Eq.(12,13) are equal, including the parameterization of CDF $F_y$ by $V(\varepsilon)$, which does not include $\varepsilon_Q$. The equivalence of $\lambda(y)$ of Eq.(12,14) applies only to one site **s** with homogeneous seismicity in its surrounding. We introduce now an expansion of this equivalence to the influence function $\lambda^*(y)$. This function describes the influence of any fixed source point **t** to the seismic hazard of all sites **s** with homogenized site conditions

$$\lambda^*(y) = \nu \int_{\mathbf{X}_E} \int_{\mathbf{s}} f_E(\mathbf{X}_E) \left(1 - F_y\left(y; E(Y(\mathbf{s})) = g(\mathbf{X}), V(Y(\mathbf{s})) = g^2(\mathbf{X}) V(\varepsilon)\right)\right) ds d\mathbf{X}_E, \; \mathbf{X} = (\mathbf{X}_E, \mathbf{s}, \mathbf{t}). \tag{14}$$

This integral is basically equivalent to the integral of Eq.(12) and the principle of area-equivalence also applies. Area-equivalent GMRs result in equal influence functions. Therein, homogeneity of seismicity is not required for Eq.(14); $f_E$ and $\nu$ can be source point specific. Time dependence is also possible.

What is the consequence? For an actual and area-equivalent GMR is $g_{actual}(\mathbf{X}) \neq g_{equivalent}(\mathbf{X})$ for almost all $\mathbf{X}$, what includes a local bias. An example is given in Fig.3b. But $\varepsilon_Q = g_{actual}(\mathbf{X})/g_{equivalent}(\mathbf{X})$ of Eq.(7) is not an actual random component and is not considered in Eq.(12-14). If $\varepsilon_Q$ resp. $\xi_Q$ are interpreted as an actual random component and the estimated residual variance from the regression analysis for the GMR is directly applied to the GMR in PSHA, then we overestimate the entire random component (residual) $\varepsilon$ resp. $\xi$ and the variances $V(Y(\mathbf{s}))$ and by this the influence function $\lambda^*(y)$ resp. the entire influence of each source point **t** on the AEFs $\lambda(y)$ of all sites. The hazard estimation of all sites is systematically over estimated in this way. This does not exclude the possibility of local underestimation of $\lambda(y)$ according to Fig.3b. The only exception of the systematic overestimation is the case if all random components $\varepsilon_Q, \varepsilon_0$ and $\varepsilon_E$ have no influence (s. above).

Non-mathematicians can numerically check these results. Beside this, we state that GMRs could be random functions because there is no proof that all events need to have equal parameters **θ** in Eq. (2-5).

### 4.2 A model of anisotropic GMR for a point source

An anisotropic GMR can be simply formulated for a point source model according to the intercept theory (Fig.3a) by a unit-isoline which includes area $\pi$ equal to the unit-circle of angle functions. The radius function $d_{unit}(\varphi)$ determines the unit-isoline with azimuth $\varphi$ of local polar coordinates with origin $\mathbf{t}$. The distance $d$ in $r^2 = h^2 + d^2$ is replaced by $d^* = d/d_{unit}(\varphi)$. An example is pictured in Fig.3b. Different unit-isolines can obviously be combined by the sum $d^2_{unit}(\varphi) = \Sigma_i a_i\, d^2_{i,unit}(\varphi)$ with weighting $0 \leq a_i \leq 1$ and $\Sigma_i a_i = 1$. We do not discuss any physical interpretation because our focus is on statistical modeling and physics is also not discussed in other anisotropic models (Enescu and Enescu 2007, Sørensen et al. 2010).

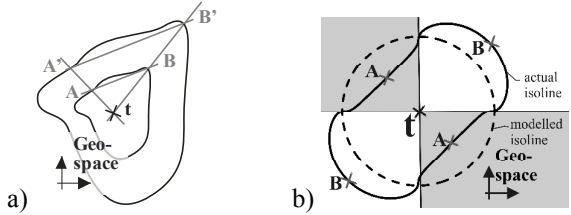

Fig.3: Construction of anisotropic GMR by an unit-isoline: a) the intercept theorem with relation between distances $(\mathbf{A'}\text{-}\mathbf{A})/(\mathbf{A}\text{-}\mathbf{t}) = (\mathbf{A'}\text{-}\mathbf{B'})/(\mathbf{A}\text{-}\mathbf{B})$, b) example of an unit-isoline with $d_{unit}(\varphi) = 0.96 + 0.352|sin(2\varphi)|^{1.5}/sin(2\varphi)$ (gray regions – overestimation, white regions – underestimation, A and B for Fig.4a)

### 4.3 Examples of area-equivalent GMRs

We illustrate the action of misinterpretation of $\varepsilon_Q$ as an actual random component in example I. We apply the unit-isoline of Fig.3b and we set $\theta_3 = 1$ and $\theta_2 = 0$ of a GMR according to Eq.(5) (s. Fig.4a). The parameters are $\theta_0 = \theta_1 = \theta_5 = 0$ because they are not relevant here. Furthermore, we fix $h = 10$km and simulate for a fixed site in the center of a source region with homogeneous seismicity as described in appendix A3. There is no actual random component in the actual GMR because $V(\xi) = 0$. We have plotted $ln(Y)$ in relation to distance $r$ in Fig.4b with the regression function for the isotropic, circular GMR. The estimated parameters are almost equal to the actuals. If the observed residuals are interpreted as actual random components, then the residual variance is overestimated with $V(\xi) = 0.10$. An interesting aspect constitutes the distance dependency of $\xi_Q$ in Fig.2b: it increases with increasing distance. But we can also construct an example II of area-equivalence with decreasing variance using the Joyner-Boore distance. In Fig.4c, an actual and an estimated vertical projection of the rupture is pictured. The shape of both projections and the included area is equal, only the azimuth is different. Obviously, the actual and the modeled Joyner-Boore distance are area-equivalent for the same GMR. But there is a component $\varepsilon_Q$ resp. $\xi_Q$. We simulate again observations without actual random components and show these in Fig. 4d. $V(\xi_Q)$ decreases with increasing, large distances.

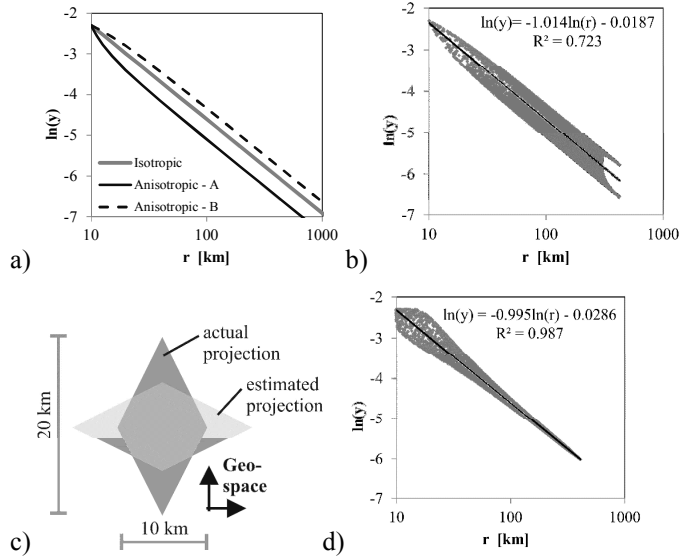

Fig.4: Area-equivalent GMR $g^*(X)$: a) isotropic and anisotropic variant of example *I* (direction A and B according to Fig.3b), b) estimation of isotropic (circular) $g^*(X)$ for *a)* with a Monte Carlo simulated sample, c) vertical projection of example *II*, d) estimated $g^*(X)$ for estimated projection and simulated sample

# 5 Numerical studies

## 5.1 The influence of the distribution type

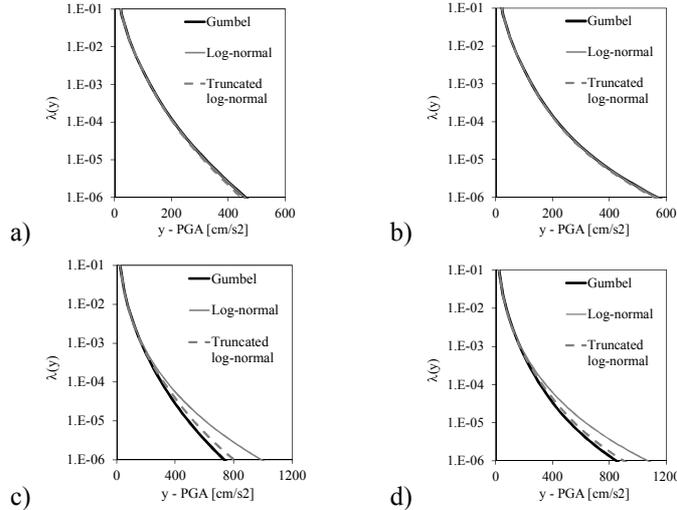

Fig.5: AEF for different distribution types: a) $V(\xi)=0.15^2$ and $m_{max}=7$, b) $V(\xi)=0.15^2$ and $m_{max}=9$, c) $V(\xi)=0.3^2$ and $m_{max}=7$, d) $V(\xi)=0.3^2$ and $m_{max}=9$ ($\xi$ here for $log_{10}$)

We numerically investigate the influence of the type of CDF $F_y$, which is determined by the distribution of $\varepsilon$ resp. $\xi$, on the hazard curve for equivalent residual variance. For this purpose, we use again the constructed situation of seismicity according to appendix A3 with fixed depth $h$=10km and consider different upper magnitudes $m_{max}$=7 and 9. Additionally, we consider different variances $V(\xi)$=0.15² and 0.3² for $log_{10}$, which are typical for previous GMRs. The applied GMR is $g^*(X)=0.5m-ln(r)-0.002r+4.7$. We consider different distributions of random component $\varepsilon$: Gumbel, the log-normal and truncated log-normal distribution. The latter has an upper and lower bound at three times its standard variation. The computed AEFs are shown in Fig.5. We note that the influence of the distribution type depends on the maximum magnitude, the residual variance and the range of *y*. The hazard of rare events is largest for the log-normal distribution with high variance. Of course, further seismicity parameters influence the contribution of the distribution model and the distribution model has no influence in case of unbounded exponentially distributed magnitudes (Sec. 4.1).

## 5.2    *An example of area-equivalent GMRs in a PSHA*

We research the influence of the misinterpretation of $\varepsilon_Q$ of example I (Sec.3.5, Fig.3b) on the PSHA. The GMR is *g\*(X)=0.5m-ln(r)+4.7*. The considered site **s** again is the centre of the quadratic source region of uniform seismicity with $m_{max}$=8; for further details see appendix A3. We compute the AEF for an isotropic (circular) GMR and the anisotropic one, both without a random component resp. residual. In the third variant, we consider the isotropic GMR and consider a normally distributed random component $\xi$ with $V(\xi)$=0.10 according to the estimation of example I in Sec.3.5. The results are illustrated in Fig.6.

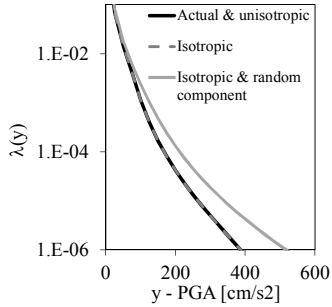

Fig.6: Hazard curves for the example of misinterpreted differences of GMR (s. Fig.4a and b)

As expected according to the theory given in Sec. 4.1, the area-equivalent GMRs without random component result in equivalent hazard and the misinterpretation of the differences as random component results in the overestimated hazard.

## 5.3    *The obscuration of a Gumbel distributed random component $\varepsilon_0$*

The Gumbel distribution of random component $\varepsilon_0$ (Eq.(7a)) could be hidden. If we observe/estimate the product $\varepsilon_0\varepsilon_S$ ($\varepsilon_S$ as unknown site effect, acting as random variable according to Joyner and Boore 1993), then we cannot test the assumption of log-normal distribution for each component. But the product could be distributed similarly to a log-normal distribution. An example: $\varepsilon_0$ is Gumbel distributed with $E(\varepsilon_0)$=1 and $V(\varepsilon_0)$=0.199, $\varepsilon_s$ is Beta distributed (s. Eq.(A3)) with $E(\varepsilon_s)$=1 and $V(\varepsilon_s)$=0.091 and with bound 0.5≤$\varepsilon_s$≤2.8. The product $\varepsilon_0\varepsilon_S$ has a mixed distribution as shown in Fig.7a. It looks very similar to a log-normal distribution. However, their tails in Fig.7b differ considerably. The tail is important for PSHA according to the studies of Restrepo-Velez and Bommer (2003) and Strasser et al. (2008), and it is a specially studied object in extreme value statistics (Leadbetter et al. 1983). Therein, the tail of a distribution should be modeled by the generalized Pareto distribution. Huyse et al. (2010) has already modeled the tail for the residuals of a ground motion using a generalized Pareto distribution.

Furthermore, we research in detail the possibility of hidden distributions for a GMR $Y=\varepsilon_0\varepsilon_S exp(0.7m-ln(r)+\theta_{s,1}x_{s,1}+\theta_{s,2}x_{s,2}+\theta_{s,1}x_{s,2})$ with a site parameter $exp(\theta_{s,i})$=0.8, 1.1 and 1.2 and indicator variables $x_{s,i}$ for three site types. We Monte Carlo generate a large sample $(ln(Y),m,ln(r),x)$ with uniform distributed magnitude with $4\leq m\leq7.5$, uniform distributed $ln(r)$ with $ln(5)\leq ln(r)\leq ln(200)$ and an occurrence probability of 1/3 for each site type. Then we carry out a LS regression using this sample and analyze the estimated residuals $\xi$ to be normally distributed. We have done this for a sample size of n=1500. The estimated residual variance is $V(\xi)=0.293$. The q-q normal plot of the residuals is shown in Fig.7c. Similar plots (s. Bommer et al. 2004, Fig.2) have been interpreted as a proof for a normally distributed $\xi$ and the wrong KS test would also indicate a log-normal distribution. Only the AD test would reject the false assumption. However, also this correct test would often accept the false distribution model. We state that the actual distribution of $\varepsilon$ resp. $\xi$ could be hidden.

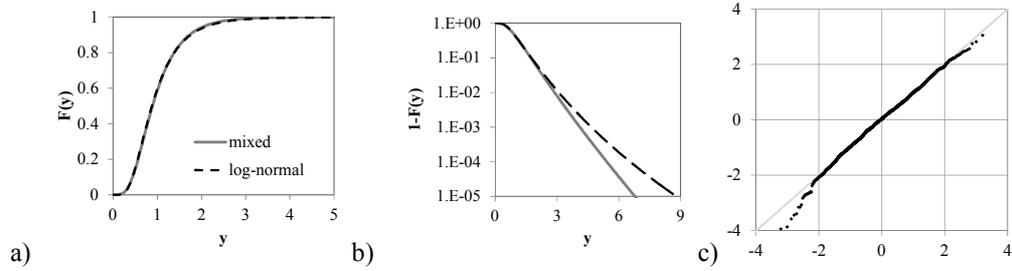

Fig.7: Possibility of hidden Gumbel distribution: a) mixed distribution of beta and Gumbel distribution and log-normal distribution with $E(ln(Y))$=-0.13 and $V(ln(Y))$=0.292, b) survival functions of *a)*, c) q-q plot

### 5.4 The influence of the different effects on the estimation of GMR and PSHA

Now we research the effects of the combination of misinterpreted random component $\varepsilon_Q$, incorrect distribution assumption for $F_y$ and measurement errors in the predictors on the PSHA using examples of anisotropic GMRs with point sources. For this purpose, we assume again the constructed situation of a site **s** and surrounding homogeneous seismicity according to appendix A3. The magnitude is upper bounded here by $m_{max}$=8. The seismicity parameters are precisely known for the PSHA but the parameters of the GMRs are estimated. For the latter, regression models are estimated for Monte Carlo simulated samples of $(Y, m_{measured}, r_{measured})$. Therein, the actual hypocenter depth $h$ is fixed and the accidental distance $d$ to the point source is beta distributed; the related parameter depends partly on the simulated beta distributed magnitude (for details s. Sec. A3).

The measurement errors of depth $h$ and distance $d$ are also Monte Carlo simulated; $h_{measured}$ is a log-normally distributed random variable; the measured distance is $d_{measured} = \left| d_{actual} + d_{error} \right|$ [km] with normally distributed $d_{error}$ with a standard deviation of 5km and an expectation of 0. A seismological epicenter or point of maximum energy could be estimated more precisely, but the seismological epicenter can differ from the epicenter of the GMR – the point of maximum $g(\mathbf{X})$ resp. $g^*(\mathbf{X})$. We also assume a normally distributed measurement error for the magnitude with a standard deviation of 0.15 and 0.25. This is plausible according to our discussion in Sec.2.2 and the magnitude errors in the PEER database (2013, *NGA_Flatfile_2005Version.xls*). We simulate 500 pairs of $(m_{measured}, r_{measured})$ for each sample using this procedure. Examples are illustrated in Fig.8. These are conceivable possibilities according to actual samples (e.g. Ambraseys and Simpson 1996, Ambraseys et al. 1996, Spudich et al. 1999, Atkinson 2004, Kalkan and Gülkan 2004, Massa et al. 2008).

The related ground motion intensity $Y = \varepsilon_S \varepsilon_0 g(\mathbf{X})$ is computed by the actual pair $(m.r)$ and the defined GMR. It is formulated by $g^*(\mathbf{X})$ and $V(\xi_0)$ of Eq.(3) and is transformed to $g(\mathbf{X})$ and $V(\varepsilon_0)$ by Eq.(4). Its relevant parameters are listed in Tab.2. The individual random component $\varepsilon_0$ is Gumbel distributed (Eq.(16b)). The site-specific random component $\varepsilon_S$ is beta distributed with expectation $E(\varepsilon_S)=1$ with a small share to $\varepsilon$ (s. Tab.2, rows 9 and 10). Anisotropy is considered by an elliptic unit-isoline, (s. Sec.4.2 and Fig.A3b). The actual random component (residuals) $\varepsilon$ is not log-normally distributed resp. $\xi$ is not normally distributed.

We estimate a GMR with the LS regression for each sample of size $n=500$ and test the estimated residuals $\xi$ to be normally distributed using the KS test as done in previous researches. We repeat this 100 times for each researched variant. The averages of the estimated parameters are listed in Tab.2 as well as the shares of rejection of the KS test. The false normal assumption for $\xi$ is accepted in 68% to 98% of the samples. The residual variances of all four variants are overestimated according to rows 10 and 11 in Tab.2. Therein, the contribution of the magnitude error is small (raw 15). We show the GMRs $g(\mathbf{X})$ in Fig.9 with actual parameters and with the averages of the estimated parameters. They do not differ from each other very much, but there is a certain bias.

Furthermore, we compute the AEF by PSHA and for the assumed seismicity described above. We compare the influence of the actual and the estimated GMRs. We apply the averages of the estimations to the latter one. The corresponding AEFs for the constructed seismicity are depicted in Fig.10. The actual AEFs are shown for site condition $E(\varepsilon_S)=1$ and the 80% quantile of $\varepsilon_S$. This gives an impression of the small influence of the considered site effects. Furthermore, we show an AEF for the area-equivalent isotropic GMR with the actual type and variance of $F_y$. We state: The area-equivalence works well, as expected. The overestimated variance and the wrong log-normal assumption lead to an overestimation of the hazard for long return periods (reciprocal of exceedance frequency). The bias in parameter vector $\boldsymbol{\theta}$ partly compensates this overestimation. The theoretical results of Sec.4.1 are confirmed.

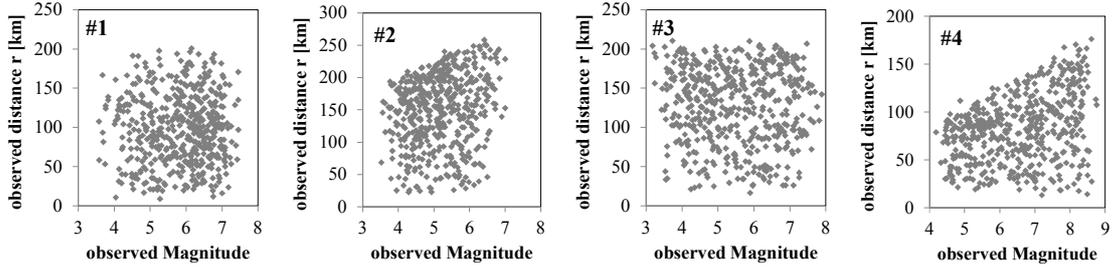

Fig.8: Examples of simulated samples

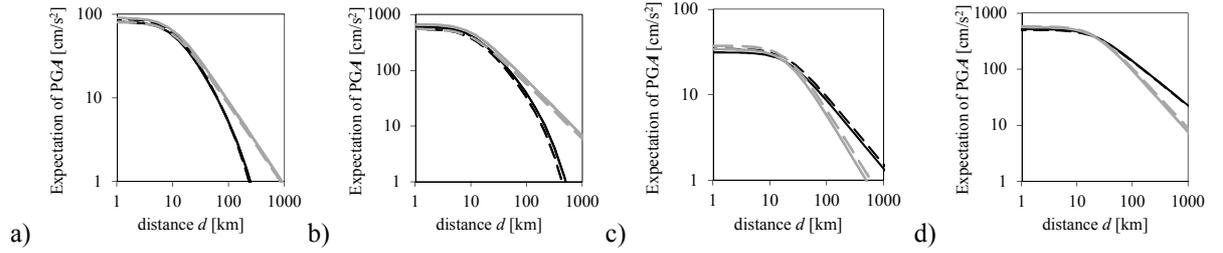

Fig.9: Actual and estimated GMRs $g(X)$ according to Tab.1, a) #1 (black line) and #2 (gray line) for $m=4$, b) as *a)* for $m=8$, c) #3 (black line) and #4 (gray line) for $m=4$, d) as *c)* for $m=8$ (full – actual, broken – estimation)

Tab.2: Investigated variants of GMRs according to Eq.(1-5) and the estimations (±standard error of the estimations; parameters $\theta_i$ are according to Eq.(5); s. also Tab.A2 and A3)

| # | Parameter | Researched variant | | | |
|---|---|---|---|---|---|
| | | **#1** | **#2** | **#3** | **#4** |
| 1 | Actual parameter $\theta_1$ | 0.5 | 0.5 | 0.7 | 0.7 |
| 2 | Average of estimated parameter $\theta_1$ | 0.4587 ±0.0324 | 0.4805 ±0.0261 | 0.6670 ±0.0149 | 0.6837 ±0.0170 |
| 3 | Actual parameter $\theta_2$ | 0.0050 | 0 | 0 | 0 |
| 4 | Average of estimated parameter $\theta_2$ | 0.0059 ±0.0007 | 0 defined | 0 defined | 0 defined |
| 5 | Actual parameter $\theta_3$ | 1 | 1 | 0.8 | 1.1 |
| 6 | Average of estimated parameter $\theta_3$ | 1 defined | 0.9860 ±0.0436 | 0.7960 ±0.0368 | 1.0670 ±0.0402 |
| 7 | Actual parameter $\theta_0$ | 4.7000 | 4.7500 | 3.000 | 4.000 |
| 8 | Average of estimated parameter $\theta_0$ | 4.7838 ±0.2114 | 4.70155 ±0.2313 | 3.0987 ±0.2030 | 3,8759 ±0.1687 |
| 9 | Actual $V(\xi_a)$ | 0.1000 | 0.1100 | 0.0800 | 0.0500 |
| 10 | Actual $V(\xi)=V(\xi_a)+V(\xi_r)$ | 0.1133 | 0.1179 | 0.0879 | 0.0633 |
| 11 | Average of estimated $V(\xi)$ | 0.4303 ±0.0226 | 0.4142 ±0.0319 | 0.3185 ±0.0232 | 0.4164 ±0.0268 |
| 12 | Actual depth $H$ [km] | 10 | 15 | 20 | 20 |
| 13 | Error of $H_{obs}$ [km] | 3 | 5 | 5 | 5 |
| 14 | Max radius of unit ellipse | 1.7 | 1.5 | 1.5 | 1.5 |
| 15 | Error of $M_{obs}$ (resulting bias of $V(\bar{\xi})$) | 0.25 (0.0142) | 0.15 (0.0055) | 0.25 (0.0288) | 0.15 (0.0109) |
| 16 | Min. of site effect $\varepsilon_s$ | 0.7857 | 0.80 | 0.80 | 0.7857 |
| 17 | Max. of site effect $\varepsilon_s$ | 1.2857 | 1.20 | 1.20 | 1.2857 |
| 18 | $p$ of site effect $\varepsilon_s$ (Eq.(A3)) | 1.5 | 2.0 | 2.0 | 1.5 |
| 19 | $q$ of site effect $\varepsilon_s$ (Eq.(A3)) | 2.0 | 2.0 | 2.0 | 2.0 |
| 20 | Share of accepted models (KS test) | 86% | 68% | 89% | 83% |

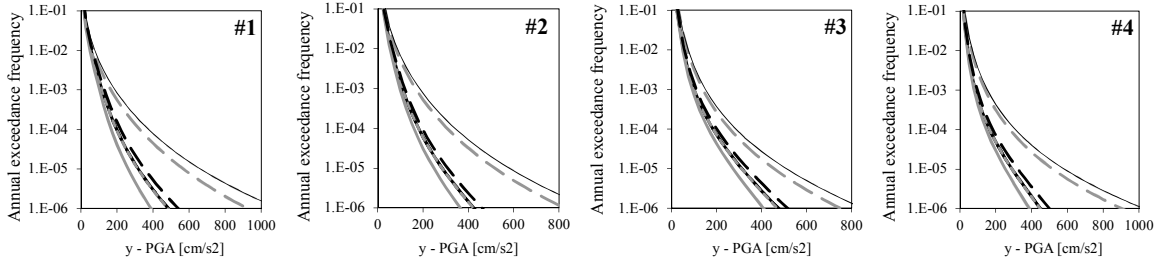

Fig.10: Estimated AEFs $\lambda(y)$ for actual and estimated parameters $\theta$, variances and distributions, # of variant right upper corner (bold, black line – actual $\theta$ and $V(\xi)$, Gumbel, anisotropic, $\varepsilon_S$=1; bold, dotted, gray line – actual $\theta$ and $V(\xi)$, Gumbel, isotropic, $\varepsilon_S$=1; bold, broken, black line – actual $\theta$ and $V(\xi)$, Gumbel, anisotropic, 80% quantile of $\varepsilon_S$; bold, dotted, gray line – actual $\theta$ and $V(\xi)$, log-normal, isotropic; bold, broken, gray line – estimated $\theta$ and $V(\xi)$, log-normal; thin, black line – actual $\theta$ and estimated $V(\xi)$, log-normal; bold, gray line – estimated $\theta$ and actual $V(\xi)$, log-normal)

# 6 Alternative estimation of GMR

## 6.1 The basic concept

We have stated in Sec.4.1 that the GMR of an earthquake is a random function in our model. That means that the GMR has event-specific parameters. In consequence, the parameters of the GMR can and should be estimated event-specific. Therein, we assume that it is practically impossible to find the "true" GMR being absolutely exact for each azimuth and with a perfect source model. Furthermore, we assume that an event-specific, area-equivalent GMR $g(\mathbf{X})$ can be formulated and estimated by a regression model, except the variances of the actual random components. The relation of the event-specific GMRs to the event magnitudes has to be researched in this concept after a number of GMRs have been estimated. This approach has already been applied by Joyner and Boore (1981): they estimated the parameter $\theta_I$ of Eq.(5) by a regression analysis of the pairs $(m, \theta_0)$, wherein $\theta_0$ of our Eq.(5) is event-specific. Event-specific random component $\varepsilon_E$ resp. $\xi_E$ would be the residual variance of such secondary regression analysis. The influence of site effects can be analyzed using a posterior analysis of the estimated residuals (Morikawa et al. 2008; negligence of non-iid). Thus, event-specific randomness of the site effects could be considered. The remaining problem is to estimate $V(\varepsilon_0)$ resp. $V(\xi_0)$ under exclusion of any influence of $\varepsilon_Q$ resp. $\xi_Q$. This should be possible by an analysis of the two horizontal components $Y_1$ and $Y_2$. The difference

$$\ln(Y_1) - \ln(Y_2) = \ln\big(g(X)\varepsilon_E\varepsilon_S\varepsilon_{01}\varepsilon_Q\big) - \ln\big(g(X)\varepsilon_E\varepsilon_S\varepsilon_{02}\varepsilon_Q\big)$$
$$= \ln(\varepsilon_{01}) - \ln(\varepsilon_{02}) = \xi_{01} - \xi_{02}, \quad E\big(\ln(\varepsilon_{01}) - \ln(\varepsilon_{02})\big) = 0 \tag{15}$$

includes only the horizontal random components $\varepsilon_{01}$ and $\varepsilon_{02}$ resp. $\xi_{01}$ and $\xi_{02}$. Therein $\varepsilon_{01}$ and $\varepsilon_{02}$ (resp. $\xi_{01}$ and $\xi_{02}$) have an equivalent distribution and they are interdependent. If we would know the dependence structure between $\varepsilon_{01}$ and $\varepsilon_{02}$ according to Mari and Kotz (2001,

Sec.4, copula), then we could estimate the distribution of $\varepsilon_{01}$ and $\varepsilon_{02}$ with the difference $ln(\varepsilon_{01})\text{-}ln(\varepsilon_{02})$ by statistical computations. Therefore, the dependence structure should be investigated in future researches. A differentiation by classes of magnitudes, regions, site conditions or something else is possible because there should be enough ground motion observations to compute a large number $ln(\varepsilon_{01})\text{-}ln(\varepsilon_{02})$. We cannot prove the functionality of our entire suggestion, but we estimate the GMRs of different earthquakes to demonstrate the potential of our approach.

### 6.2 Analysis of empirical data

We analyze the observed PGAs of nine earthquakes of the PEER strong motion database (2013, files: *NGA_Flatfile_2005Version.xls, NGA_Documentation.xls*). We select such earthquakes with a large number of records and an event center inside the cloud of strong motion stations, which should cover the entire event area (event # 136 differs slightly). Furthermore, we consider only one event from a cluster; try to consider different regions and to cover a relatively large range of the magnitude scale. The selected earthquakes are listed in Tab.3. We consider different models: a point source model with isotropic GMR, a point source model with anisotropic GMR and (if available) the source models which lead to the constructed Joyner-Boore distance, the Campbell distance, the root-mean-squared distance (RmsD) and the closest distance to the ruptured area (ClstD). Our basic formulation is $g^*(X)=\theta_0\text{-}\theta_2 ln(r)\text{-}\theta_3 r$ according to Eq.(5) with $r^2=d^2+h^2$ resp. $r^2=d^2/d^2_{unit}(\varphi)+h^2$ for the anisotropic point source. We use an eccentric circle and an ellipse (Fig.A3) to model anisotropy. Additionally, we consider different combinations of defined and estimated parameters. The depth parameter $h$ can be set by the hypocenter depth or be estimated with limit $h{\geq}0.1$km. The parameter $\theta_3$ for $ln(r)$ is estimated or set to 1; we do not consider a bound. The parameter $\theta_2$ is either set to 0 or estimated with limit $\theta_2{\geq}0$. We divide the models into groups: the constructed distances, the isotropic point source with epicenter as projected point source, the isotropic point source with estimated coordinates of the point source (start values are the epicenter coordinates) and the anisotropic point source model with estimated coordinates of the point source. For each division, we select the variant of best combination of estimated/set parameters by the smallest AIC (Rawlings et al. 1998, Sec.7) with sample size $n$ and parameter number $N$

$$AIC = \ln(\hat{V}(\xi)) + 2N/n\,.\tag{16}$$

The parameters of the best models are listed in Tab.A4. Their AICs are given in Tab.4, $V(\xi)$ in Tab.5. The constructed distances do not result in good estimations; the anisotropic point source model is frequently the best model. This may be relative because constructed distances do not exist for each event, but we consider four constructed distances and only two simple variants of anisotropy. Additionally, it may be possible that we have estimated only a local minimum of least squares for the anisotropic models, the global one could be much better. The average variance of the best models of all point source models is $V(\xi)$=0.19. This also includes the site effects, but it is nevertheless significantly smaller than the residual variance of the intra-event component of NGA. For example, Abrahamson and Silva (2008, standard derivations $s_1$ and $s_2$ of Tab.6, Eq.27,) have variances between 0.35 (m≤5) and 0.22 (m≥7). This fact indicates the advantage of our approach. Examples of the estimated GMRs are shown in Fig.11. The graphs of the GMRs look partly very individual, which is a result of event-specific parameters. This validates the approach of individual GMRs for individual earthquakes. There are also some cases with estimated depth $h$=0.1km; our defined lower bound for $h$. Reason for such poor estimations is the non-regular situation in the regression model: a parameter $h$ defines the predicting variable – distance $r$. Smith (1985) has mathematically researched the problem of irregularity for distribution functions; we do not know a similar research for regression models. However, this problem could be minimized by the Bayesian approach of parameter estimation: the seismological source estimation could provide a prior distribution for the depth parameter. Furthermore, the estimations for event #136, the Kocaeli, Turkey 1999 earthquake are poor. The parameter $\theta_3$ is <0 in some estimations. However, we do not change or remove this event.

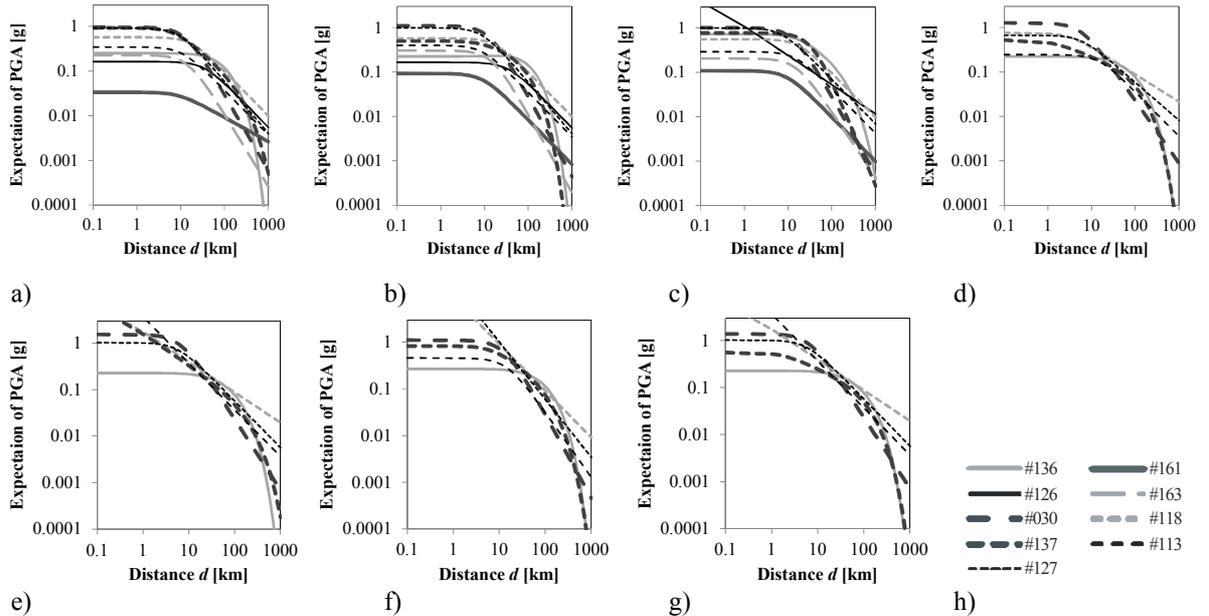

Fig.11: Estimated GMRs $g(X)$: a) iso. point source, source coord.=epicenter, b) iso. point source, estimated source coord., c) an-iso. point source, estimated source coord., d) Joyner-Boore distance, e) Campbell distance, f) RmsD, g) ClstD, h) legend

Tab.3: Analyzed earthquakes of the PEER strong motion database

| # | Earthquake | Magnitude | Latitude (Epic.) | Longitude (Epic.) | Hypocenter depth [km] | Sample size $n$ |
|---|---|---|---|---|---|---|
| 030 | San Fernando, California 1971 | 6.61 | 34.44 | -118.41 | 13 | 44 |
| 113 | Whittier Narrows-01, California 1987 | 5.99 | 34.05 | -118.08 | 14.6 | 116 |
| 118 | Loma Prieta, California 1989 | 6.93 | 37.04 | -121.88 | 17.48 | 82 |
| 126 | Big Bear-01, California 1992 | 6.46 | 34.21 | -116.83 | 13 | 73 |
| 127 | Northridge-01, California 1994 | 6.69 | 34.21 | -118.55 | 17.5 | 157 |
| 136 | Kocaeli, Turkey 1999 | 7.51 | 40.73 | 29.99 | 15 | 31 |
| 137 | Chi-Chi, Taiwan 1999 | 7.62 | 23.86 | 120.8 | 6.76 | 420 |
| 161 | Big Bear-02, California 2001 | 4.53 | 34.29 | -116.95 | 9.1 | 43 |
| 163 | Anza-02, California 2001 | 4.92 | 33.51 | -116.51 | 15.2 | 73 |

Tab.4: Best AICs of the different model approaches (bolted – absolute best, cursive – best of point sources)

| Approach | | Distance (to) | Earthquake # | | | | | | | | |
|---|---|---|---|---|---|---|---|---|---|---|---|
| | | | 127 | 113 | 137 | 118 | 030 | 163 | 126 | 161 | 136 |
| Point source | Isotrop. | Seism. Epicenter | -1.61 | -1.33 | -1.12 | *-1.31* | -1.16 | -1.24 | -1.71 | -1.23 | -1.06 |
| | | Estimated Epicenter | -1.65 | -1.32 | -1.47 | -1.29 | -1.42 | -1.21 | -1.71 | -1.37 | -1.08 |
| | Aniso. | Estimated Epicenter | ***-1.74*** | ***-1.55*** | ***-1.76*** | -1.30 | ***-1.66*** | ***-1.42*** | ***-2.02*** | ***-1.39*** | *-1.30* |
| Con-structed distance | | Joyner-Boore | -1.71 | -1.33 | -1.31 | **-1.43** | -1.45 | - | - | - | -1.34 |
| | | Campbell | -1.59 | -1.32 | -1.32 | -1.37 | -1.48 | - | - | - | **-1.34** |
| | | RmsD | -1.67 | -1.32 | -1.45 | -1.34 | -1.47 | - | - | - | -1.22 |
| | | ClstD | -1.59 | -1.32 | -1.28 | -1.37 | -1.48 | - | - | - | -1.34 |

Tab.5: Residual variances $V(\xi)$ of $g^*(\mathbf{X})$ for $ln(Y)$ of the different model approaches

| Approach | | Distance (to) | Earthquake # | | | | | | | | |
|---|---|---|---|---|---|---|---|---|---|---|---|
| | | | 127 | 113 | 137 | 118 | 030 | 163 | 126 | 161 | 136 |
| Point source | Isotrop. | Seism. Epicenter | 0.191 | 0.255 | 0.321 | 0.262 | 0.285 | 0.273 | 0.176 | 0.267 | 0.286 |
| | | Estimated Epicenter | 0.180 | 0.248 | 0.225 | 0.257 | 0.201 | 0.267 | 0.167 | 0.222 | 0.246 |
| | Aniso. | Estimated Epicenter | 0.163 | 0.195 | 0.166 | 0.253 | 0.159 | 0.206 | 0.110 | 0.217 | 0.240 |
| Con-structed distance | | Joyner-Boore | 0.177 | 0.263 | 0.266 | 0.228 | 0.214 | - | - | - | 0.231 |
| | | Campbell | 0.201 | 0.263 | 0.264 | 0.241 | 0.207 | - | - | - | 0.230 |
| | | RmsD | 0.184 | 0.262 | 0.232 | 0.254 | 0.219 | - | - | - | 0.260 |
| | | ClstD | 0.201 | 0.263 | 0.274 | 0.241 | 0.208 | - | - | - | 0.230 |

## 6.3 Area functions and site effects of the Chi-Chi earthquake

The Chi-Chi, Taiwan 1999 earthquake (#137) is the one with the largest sample size $n$=420. We can use it to compare the area function of the estimated GMR to the actual one. But we cannot compute the area function directly because the records are from stations that are not uniformly distributed; there are concentrations and thinning that have to be considered. We do it using an empirical area function wherein the integration of Eq.(9) is replaced by a discrete accumulation with

$$\hat{K}_{observed}(y) = \sum_{i=1}^{n} a_i^* \mathbf{1}(Y_i), \quad \mathbf{1}(Y_i) = 1 \ \textit{if } Y_i \geq y, \ \textit{otherwise } \mathbf{1}(Y_i) = 0 \ \text{and} \quad (17)$$

$$\hat{K}_{GMR}(y) = \sum_{i=1}^{n} a_i^* \mathbf{1}(g(\mathbf{X}_i, \hat{\boldsymbol{\theta}})), \ \mathbf{1}(g(\mathbf{X}_i, \hat{\boldsymbol{\theta}})) = 1 \ \textit{if } g(\mathbf{X}_i, \hat{\boldsymbol{\theta}}) \geq y, \ \textit{otherwise } \mathbf{1}(g(\mathbf{X}_i, \hat{\boldsymbol{\theta}})) = 0 \ . \quad (18)$$

We estimate these discrete steps $a_i^*$ by a Voronoi analysis of the stations and our estimated area functions are defined with indicator function (s. Eq.(9)). The results are shown in Fig.12. A good model should include a good accordance between Eq.(17,18) with larger differences for larger $y$ because of a larger influence of the random components. Additionally, a certain bias is conceivable for very small values of $y$ because of the effect of the truncation of the geo-space by the finite sample resp. station number in Eq.(17,18). A good agreement is detected for the point source model with estimated coordinates. Especially the anisotropic variant fits well, in contrary to the constructed distances.

The plausibility of the comparison of $\hat{K}_{observed}(y)$ and $\hat{K}_{GMR}(y)$ can be simply tested by generation of $Y = \varepsilon g(\mathbf{X}_i, \hat{\boldsymbol{\theta}})$ with the estimated GMR and Monte Carlo simulated random component $\varepsilon$. We do it for anisotropic GMR for a point source and simulate 100 times the entire sample and adopted the weighting $a_i^*$ for $\hat{K}_{observed}(y)$ by factor 0.01. We consider a log-normal and gamma distributed random component to demonstrate the generality of the approach (gamma distribution: s. Johnson et al. 1994, Sec.17). Therein is $E(\varepsilon)=1$ and $V(\varepsilon)=0.181$, what corresponds with $V(\xi)=0.166$ (s.Eq.(4)). The results are pictured in Fig.12h. The approach works, and the distribution type of $\varepsilon$ is not relevant for the medium range of PGA.

We also estimate the site effects for the Campbells GEOCODE of the PEER data (2013, *NGA_Documentation.xls*) using the expectation of the residuals (s. Tab.6).

Tab.6: Expectations of residuals of $g^*(\mathbf{X})$ and the statistical significance for different site classes

| Site class | Expectation of residuals of ln(Y) | Sample size | Standard error | Significance to be ≠0, α=5% |
|---|---|---|---|---|
| A | 0.077 | 199 | 0.022 | Yes |
| C | -0.077 | 209 | 0.033 | No |
| D | -0.310 | 3 | 0.046 | Yes |
| F | 0.190 | 9 | 0.054 | Yes |

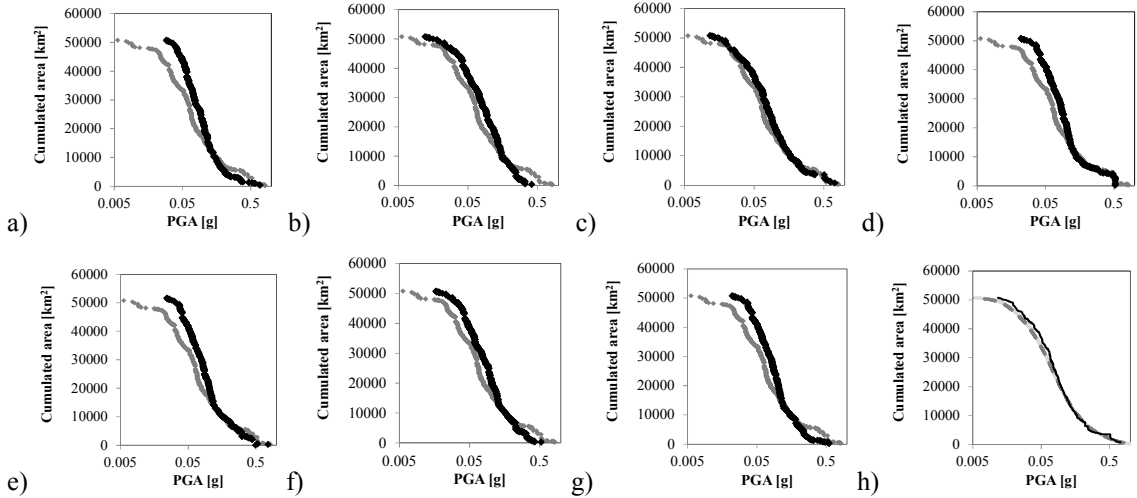

Fig.12: Area functions of different GMRs for the Chi-Chi earthquake (black – model, gray – observed): a) iso. point source with coordinates=seismo. epicenter, b) iso. point source with estimated coordinates, c) an-iso. point source with estimated source coordinates, d) Joyner-Boore distance, e) Campbell distance, f) RmsD, g) ClstD, h) validation of the comparison of the empirical area functions (bold light gray line – log-normally distributed $\varepsilon$, bold dotted dark gray line - gamma distributed $\varepsilon$, thin black line – Eq.(18))

## 6.4 Relation of specific GMRs to the magnitude

There is the need to find a relation between the event-specific GMRs and the earthquake magnitudes when the event size in the PSHA is quantified by the magnitude. We search such a relation by a statistical analysis of the relations between the parameters of the GMRs and the magnitudes. The results are shown in Fig.13a to d. Obviously, there is not a significant relation. But we follow the idea of the max-stable random fields and compute the volume of GMRs in the geo-space. We compute the volume

$$V_o = \int_S g^2(\mathbf{X}) d\mathbf{s} \qquad (19)$$

numerically for distance $d$ resp. $d^* \leq 1000$km in steps of 25m. Therein we squared the event-specific GMR because the PGA is approximately proportional to the PGV (Wald et al. 2006, section 2.5, Eq.(1.1-1.4)), the squared velocity is proportional to the energy and the energy is strongly related to the magnitude. The logarithms of these volumes have a strong statistical linear relation to the magnitude according to Fig.13e and f with a minor influence of $V(\xi)$.

Of course, the applied sample size is small, which causes an uncertainty of the result. But the magnitudes and the volumes are also only estimated and include an estimation error. Such errors rather disturb the observed relation. Therefore, the actual relation should be really strong. The relation is also not changed significantly if we eliminate event #136 with the poor estimations. Such a relation could be applied to GMRs in a PSHA. The distance parameters $\theta_2$ and $\theta_3$ in Eq.(5) could be random variables and $\theta_0$ is computed using the relation magnitude to volume. The previous magnitude parameter $\theta_1$ would not be needed anymore.

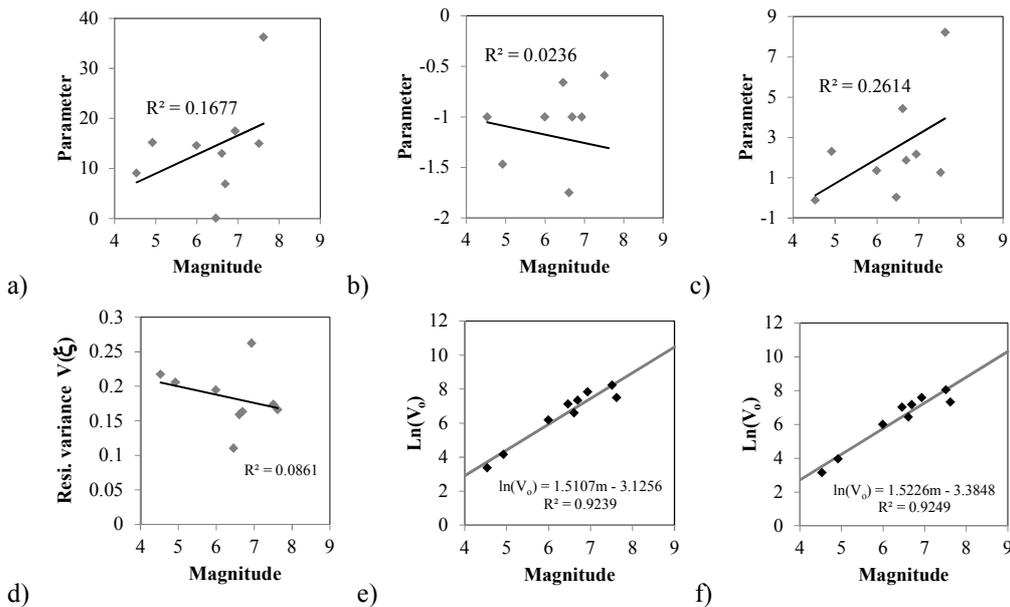

Fig.13: Relation of magnitude to the regression models with correlation coefficient $R$: a) to h, b) to $\theta_3$, c) to $\theta_0$, d) to $V(\xi)$, d) to volume according to Eq.(19), f) as e) but with $V(\xi)=0$ in $g(\mathbf{X})$ according to Eq.(4a)

# 7 Conclusion and outlook

We have discussed here important aspects of earlier approaches to GMR by a regression model and discovered in Sec.2 that many models have not been built according to the rules of statistics regarding statistical significance, model selection and test of the distribution assumption. But even if the log-normal assumption for residuals of $\varepsilon$ are tested positively, it is not because the individual component $\varepsilon_0$ according to Eq.(7) of a PGA or another maximum value should be generalized extreme value distributed according to the extreme value statistics (Sec. 3, 5). Its domain of attraction seems to be the Gumbel one, but this issue should be examined by future researches. Our major contribution is the introduction of area-equivalence of GMR for PSHA in Sec.4, which implies a distinction between an appropriate prediction of the PGA for a concrete earthquake by a conventional regression model and an appropriate GMR for the PSHA. These models need not to be equal regarding the residual variances. In contrary, the residual of the regression model for the random function GMR includes the component $\varepsilon_Q$ of Eq.(4). This may not apply as an actual random component in the GMR for the PSHA, otherwise the variance $V(\varepsilon)$ is overestimated, which leads to an overestimated hazard in the PSHA (except for one case, Sec.4.1). The possible influence of the distribution of $\varepsilon$ and the misinterpretation of $\varepsilon_Q$ have been researched in Sec.5. The overestimation of hazard can be remarkable. Our numerical studies consider a broader constellation of parameters and distribution of predictors than the numerical studies of Joyner and Boore (1993) and Chen and Tsai (2002) about estimation procedures for GMRs. Nevertheless, the benefit is limited. More extensive numerical studies with different sample sizes would be needed to quantify more exactly the bias in the PSHA. Independent of this fact, we have suggested an estimation concept for GMRs in PSHA in the last section, including the independent estimation of the parameters of the individual random component $\varepsilon_0$. We stated that the dependence structure of the horizontal components has to be researched in the future to apply this concept. However, we were able to show that the event-specific modeling of GMR leads to smaller variances $V(\xi)$ than earlier models. Therein the anisotropic point source approach results in the best regression models, while the constructed distances (e.g. Joyner-Boore) do not work well. The relation between the event magnitude and the GMRs is given by the integration of $g^2(\mathbf{X})$ over the geo-space. Details of this relation and its consideration in PSHA should be studied in further researches. Beside this, the empirical area functions for the Chi-Chi, Taiwan 1999 earthquake confirm that the anisotropic point source approach works well. Nevertheless, we also suggest developing a detailed theory of this geo-statistical approach in the future. This also applies to the estimation of point source coordinates and depth by a regression analysis. Further statistical methods like Bayes estimation, local regression or kernel regression could provide better estimations; and the models of extreme value statistics for the distribution tails could improve the GMR in PSHA. A large challenge for future researches is also the discovering, estimation and/or examination of the distribution of every single random component of the GMR.

## Acknowledgement

We would expressly like to thank Jürg Hüsler and Zakhar Kabluchko for their explanations of details of the extreme value theory and statistics.

## References

Abrahamson NA (1988) Statistical properties of peak ground motion accelerations recorded by the SMART 1 array. Bulletin of the Seismological Society of America 78: 26-41

Abrahamson NA, Youngs RR (1992). A stable algorithm for regression analyses using the random effects model, Bull. Seism. Soc. Am. 82, 505–510

Abrahamson N, Atkinson G, Boore D, Bozorgnia Y et al. (2008) Comparisons of the NGA ground-motion relations. Earthquake Spectra 24: 45–66

Abrahamson N, Silva W (2008) Summary of the Abrahamson & Silva NGA Ground-Motion Relations. Earthquake Spectra 24: 67–97

Abrahamson NA, Birkhauser P, Koller M et al.(2002) PEGASOS - A comprehensive probabilistic seismic hazard assessment for nuclear power plants in Switzerland. Proceedings of the Twelfth European Conference on Earthquake Engineering, Paper on 633, London

Al Atik L, Abrahamson N, Bommer JJ, Scherbaum F et al. (2010) The variability of ground-motion prediction models and its components. Seismological Research Letters 81: 794-801.

Ambraseys NN, Bommer J (1991) The attenuation of ground accelerations in Europe. Earthquake Engineering and Structural Dynamics 20: 1179–1202

Ambraseys NN, Simpson KA (1996) Prediction of vertical response spectra in Europe. Earthquake Engineering and Structural Dynamics 25: 401-412

Ambraseys NN, Simpson KA, Bommer JJ (1996) Prediction of horizontal response spectra in Europe. Earthquake Engineering and Structural Dynamics 25: 401-412

Ambraseys NN, Douglas J, Sarma SK (2005) Equations for the estimation of strong ground motions from shallow crustal earthquakes using data from Europe and the Middle East: Horizontal peak ground acceleration and spectral acceleration. Bulletion of Earthquake Engineering 3:1-53

Anderson JG, Uchiyama Y (2011) A methodology to improve ground-motion prediction equations by including path corrections. Bulletin of the Seismological Society of America 101: 1822–1846

Atkinson GM, Boore DM (1995) Ground-motion relations for Eastern North America. Bulletin of the Seismological Society of America 85: 17-30

Atkinson GM (2004) Empirical attenuation of ground-motion spectral amplitudes in southeastern Canada and the northeastern United States. Bulletin of the Seismological Society of America 94: 1079-1095

Atkinson GM (2006) Single Station Sigma. Bulletin of the Seismological Society of America 96: 446-445

Beirlant J, Goegebeur Y, Teugels J, Segers J (2004) Statistics of extremes: theory and applications. Wiley Series in Probability and Statistics, Wiley & Sons Chichester

Billingsley P (1995) Probability and measure. Wiles Series in Probability and mathematical statistics. Wiley & Sons, USA.

Beyer K, Bommer JJ (2007) Relationships between Median Values and between Aleatory Variabilities for Different Definitions of the Horizontal Component of Motion. Bull. Seism. Soc. Am. 96, No. 4A, pp. 1512–1522

Bommer JJ, Abrahamson NA Strasser FO et al. (2004) The challenge of defining the upper limits on earthquake ground motions. Seismological Research Letters 75(1), 82-95

Bommer JJ, Abrahamson A (2006) Why do modern probabilistic seismic hazard analyses often lead to increased hazard estimates? Bulletin of the Seismological Society of America 96: 1967-1977

Bommer JJ, Stafford PJ, Alarcón JE, Akkar S (2007) The Influence of Magnitude Range on Empirical Ground-Motion Prediction. Bull. Seism. Soc. Am. 97, No. 6, pp. 2152–2170

Boore DM, Atkinson GM (2007) Boore-Atkinson NGA Ground Motion Relations for the Geometric Mean Horizontal Component of Peak and Spectral Ground Motion Parameters. PEER Report 2007/01, Pacific Earthquake Engineering Research Center, College of Engineering, University of California, Berkeley

Campbell KW (1981) Near-source attenuation of peak horizontal acceleration. Bulletin of the Seismological Society of America 71: 2039-2070

Campbell KW (1993) Empirical prediction of near-source ground motion from large earthquakes. In: Proceedings of the International Workshop on Earthquake Hazard and Large Dams in the Himalaya. Indian National Trust for Art and Cultural Heritage, New Delhi, India

Campbell K, Bozorgnia Y (2008) NGA Ground motion model for the geometric mean horizontal component of PGA, PGV, PGD and 5% Damped linear elastic response spectra for periods ranging from 0.01 to 10 s. Earthquake Spectra 24: 139–171

Chang T, Cotton E J, Anglier J (2001) Seismic attenuation and peak ground acceleration in Taiwan, Bulletin of the Seismological Society of America 91, 1,229-1,246

Chiou BS-J, Youngs RR (2008) NGA Model for Average Horizontal Component of Peak Ground Motion and Response Spectra. PEER Report 2008/09, Pacific Engineering Research Center College of Engineering, University of California, Berkeley

Castellaro S, Mulargia F, Kagan YY (2006) Regression problems for magnitudes. Geophys. J. Int. 165: 913-930

Chen Y-H, Tsai CCP (2002) A New Method for Estimation of the Attenuation Relationship with Variance Components. Bulletin of the Seismological Society of America 92: 1984–1991

Cheng C-L, van Ness JW (1999) Statistical Regression with Measurement Error. Kendall's Library of statistics, 6, Arnold, London

Coles S (2001) An introduction to statistical modeling of extreme values. Springer, London.

Cornell CA (1968) Engineering seismic risk analysis. Bulletin of the Seismological Society of America 58: 1583-1606

Cosentino P, Ficarra V, Luzio D (1977) Truncated exponential frequency-magnitude relationship in earthquake statistics. Bulletin of the Seismological Society of America 67: 1615-1623

Crouse C B, McGuire J W (1996) Site response studies for purpose of revising NEHRP seismic provisions. Earthquake Spectra 12: 407–439

de Haan L, Ferreira A (2006) Extreme value theory. Springer, New York

D'Augustino RB, Stephens MA (Editors, 1986) Goodness-of-Fit Techniques. statistics: textbooks and monographs, Vol. 68, Marcel Dekker, New York

David J. Wald, Bruce C. Worden, Vincent Quitoriano, and Kris L. Pankow (2006) ShakeMap® Maunual. Advanced national seismic system USGS (http://pubs.usgs.gov/tm/2005/12A01/)

Douglas J, Smit PM (2001) How accurate can strong ground motion attenuation relations be? Bulletin of the Seismological Society of America 91; 1917-1923

Douglas J (2001) A comprehensive worldwide summary of strong-motion attenuation relationships for peak ground acceleration and spectral ordinates (1969 to 2000). ESEE Report 01-1. Department of Civil and Environmental Engineering, Imperial College, London (http://nisee.berkeley.edu/library/douglas/ESEE01-1.pdf)

Douglas J (2002) Errata of and additions to ESEE Report No. 01-1: 'A comprehensive worldwide summary of strong-motion attenuation relationships for peak ground acceleration and spectral ordinates (1969 to 2000)'. Dept. Report, Imperial College of Science, Technology and Medicine Department of Civil & Environmental Engineering, London (http://nisee.berkeley.edu/library/douglas/douglas2002.pdf)

Douglas J (2003) Earthquake ground motion estimation using strong-motion records: A review of equations for the estimation of peak ground acceleration and response spectral ordinates. Earth-Science Reviews 61, 43–104

Dupuis DJ, Flemming JM (2006) Modeling peak acceleration from earthquakes. Earthquake Engineering and Structural Mechanics 35: 969-987

Efron B (1979) Bootstrap Methods: Another Look at the Jackknife. The Annals of statistics 7: 1-26

Enescu D, Enescu BD (2007) A procedure for assessing seismic hazard generated by Vrancea earthquakes and its application. III. Method for developing isoseismal maps and isoacceleration maps. Application. Romania reports in Physics: 59, 121-145

Falk M, Hüsler J, Reiss R-D (2011) Laws of small numbers: extremes and rare events. 3rd Ed. Birkhäuser, Basel.

Fisher RA, Tippett LHC (1928) Limiting forms of the frequency distributions of largest or smallest member of a sample. Proc. Cambridge Philos. Soc. 24: 180-190

Giardini D (1984) Systematic analysis of deep seismicity: 200 centroid-moment tensor solutions for earthquakes between 1977 and 1980, Geophys. J. R. astr. Soc. 77, 883-914

Gnedenko BV (1943) Sur la distribution limite du terme d'une série aléatoire. Ann Math. 44: 423-453

Hüsler J, Li D, Raschke M. (2011) Estimation for the generalized Pareto distribution using maximum likelihood and goodness-of-fit. Communication in Statistics – Theory and Methods 40: 2500 – 2510.

Huyse L, Chen R, Stamatakos AJ (2010) Application of Generalized Pareto Distribution to Constrain Uncertainty in Peak Ground Accelerations. Bulletin of the Seismological Society of America 100(1): 87-101

Idriss IM (2007) Empirical model for estimating the average horizontal values of pseudo-absolute spectral acceleratins generated by crustal earthquakes. Vol.1. Interim Report Issued for USGS Review, PEER (http://peer.berkeley.edu/ngawest/nga_models.html)

Johnson NL, Kotz S, Balakrishnan N (1994) Continuous univariate distributions – Vol.I. 2nd ed., Wiley, New York

Johnson NL, Kotz S, Balakrishnan N (1995) Continuous univariate distributions – Vol.II. 2nd ed., Wiley, New York

Joyner WB and Boore DM (1981) Peak horizontal acceleration and velocity from strong-motion records including records from the 1979 imperial valley, California, earthquake. Bull. Seism. Soc. Am. No. 6, pp. 2011-2038.

Joyner WB and Boore DM (1993) Methods for regression analysis of strong motion data. Bull. Seism. Soc. Am. 83, No. 2, pp. 469-487

Kabluchko Z, Schlather M, de Haan L (2009) Stationary max-stable random fields associated to negative definite functions. The Annals of Probability 37( 5): 2042–2065.

Kaklamanos J, Baise LG (2010) Model validation of recent ground motion prediction relations for shallow crustal earthquakes in active tectonic regions, in Procceedings: 5th International Conference on Recent Advances in Geotechnical Earthquake Engineering and Soil Dynamics, May 24-29, 2010, San Diego, California

Kalkan E, Gülkan P (2004) Empirical attenuation equations for vertical ground motion in Turkey. Earthquake Spectra 20: 853–882

Landry L, Lepage Y (1992) Empirical behavior of some tests for normality. Communications in statistics – Simulation and Computation 21: 971-999

Leadbetter MR, Lindgren G, Rootzen H (1983) Extremes and related properties of random sequences and processes. Springer Series in statistics, Springer, New York, Heidelberg, Berlin

Lin PS, Chiou B, Abrahamson N, Walling M (2011) Repeatable Source, Site, and Path Effects on the Standard Deviation for Empirical Ground-Motion Prediction Models. Bulletin of the Seismological Society of America 101: 2281–2295

Lindsey JK (1996) Parametric statistical inference. Oxford science publications, Oxford university press, Oxford

Mari DD, Kotz S (2001) Correlation and dependence. Imperial College Press, London.

Massa M, Morasca P, Moratto L et al. (2008) Empirical Ground-Motion Prediction equations for Northern Italy using weak- and strong-motion amplitudes, frequency content, and duration parameters. Bulletin of the Seismological Society of America 98: 1319-1342

Molas GL, Yamazaki F (1995) Attenuation of Earthquake Ground Motion in Japan Including Deep Focus Events. Bulletin of the Seismological Society of America 85: 1343-1358

Monguilner C A, Ponti N, Pavoni S B et al. (2000) Statistical characterization of the response spectra in the Argentine Republic. In: *Proceedings of 12th World Conference on Earthquake Engineering*. Paper no. 1825

Montgomery CM, Peck EA, Vining GG (2006) Introduction to linear regression analysis. Wiley and Sons, Hoboken

Morikawa N, Kanno T, Narita A et al. (2008) Strong motion uncertainty determined from observed records by dense network in Japan. J Seismol 12: 529-546

McGuire RK (1977) Seismic design spectra and mapping procedures using hazard analysis based directly on oscillator response. Earthquake Engineering and Structural Dynamics 5: 211-234

McGuire RK (1995) Probabilistic Seismic Hazard Analysis and Design Earthquakes: Closing the Loop. Bulletin of the Seismological Society of America 85**:** 1275-1284

PEER Strong Motion Database (2000) http://peer.berkeley.edu/smcat/. accessed December 2010

PEER Strong Motion Database (2013) http://peer.berkeley.edu/peer_ground_motion_database. accessed March 2013

Quenouille M H (1956) Notes on bias in estimation. Biometrika 43: 353-60

Raschke M, Thürmer K (2008) Defizite der Modellselektion in der Hochwasserstatistik (German, Shortcomings of model selection in flood statistics) Wasser und Abfall 10 (12): 43-48.

Raschke M (2009) The Biased Transformation and Its Application in Goodness-of-Fit Tests for the Beta and Gamma Distribution. Communication in statistics – Simulation and Computation 38: 1870-1890

Raschke M (2011) Inference for the truncated exponential distribution. Stochastic Environmental Research and Risk Assessment. DOI: 10.1007/s00477-011-0458-8

Raschke M (2012) Möglichkeiten der mathematischen Statistik zur Schätzung der Hochwasserwahrscheinlichkeit (German, Possibilities of mathematical statistics to estimate flood probability). Wasser und Abfall 14(6): 49-53, (cms.springerprofessional.de/journals/JOU=35152/VOL=2012.14/ISU=6/ART=193/BodyRef/PDF/35152_2 012_Article_193.pdf).

Raschke M (2013) Parameter estimation for the tail distribution of a random sequence. Communication in statistics – Simulation and Computation 42: 1013-1043.

Rawlings JO, Pantula SG, Dickey DA(1998) Applied Regression Analysis: A Research Tool. 2$^{nd}$ Ed. Springer: New York
  (http://web.nchu.edu.tw/~numerical/course992/ra/Applied_Regression_Analysis_A_Research_Tool.pdf)
Restrepo-Velez LF, Bommer JJ (2003) An exploration of the nature of the scatter in ground-motion prediction equations and the implications for seismic hazard assessment. Journal for Earthquake Engineering 7: 171-199
Rhoades DA (1997) Estimation of Attenuation Relations for Strong-Motion Data Allowing for Individual Earthquake Magnitude Uncertainties. Bulletin of the Seismological Society of America 87, 1674-1678
Sadigh K, Chang C-Y, Egan JA et al. (1997) Attenuation relationships for shallow crustal earthquakes based on California strong motion data. Seismological Research Letters 68, 180–189
Schlather M (2002) Models for Stationary Max-Stable Random Fields. EXTREMES : 33-44.
Smith RL (1985) Maximum likelihood estimation in a class of nonregular cases. Biometrika 72:67–90
Scherbaum F, Cotton F, Smit P (2004) On the use of response spectral-reference data for selection and ranking of ground-motion models for seismic-hazard analysis in regions of moderate seismicity: The case of rock motion. Bulletin of the Seismological Society of America 94: 2164-2185
Sørensen, M.; Stromeyer, D.; Grünthal, G. (2010) A macroseismic intensity prediction equation for intermediate depth earthquakes in the Vrancea region, Romania. Soil Dynamics and Earthquake Engineering, 30, 11, 1268-1278.
Spudich P, Joyner WB, Lindh AG et al. (1999) SEA99: A revised ground motion prediction relation for use in extensional tectonic regimes. Bulletin of the Seismological Society of America 89: 1156-1170
Stafford PJ, · Strasser ·FO, Bommer JJ (2008) An evaluation of the applicability of the NGA models to ground-motion prediction in the Euro-Mediterranean region. Bull Earthquake Eng (2008) 6:149–177
Stephens MA (1986) Test based on EDF statistics. in D'Augustino, RB, Stephens, MA (Editors) Goodness-of-Fit Techniques. statistics: textbooks and monographs, Vol. 68, Marcel Dekker, New York
Stepp JC,Wong I, Whitney J  et al. (2001) Probabilistic seismic hazard analyses for ground motions and fault displacements at Yucca Mountain, Nevada, Earthquake Spectra 17(1), 113–151
Strasser FO, Bommer JJ, Abrahamson NA (2008) Truncation of the distribution of ground-motion residuals. Journal of Seismology 12(1), 79-105
Strasser FO, Abrahamson NA, Bommer JJ (2009) Sigma: Issues, insights and challenges. Seism. Res. Lett. 80: 40-56
Stromeyer D, Grünthal G, Wahlström R (2004) Chi-square Maximum Likelihood Regression for Seismic Strength Parameter Relations, and their Uncertainties, with Applications to an Mw based Earthquake Catalogue for Central, Northern and Northwestern Europe. Journal of Seismology 8:143-153
Upton G, Cook I (2008) A dictionary of statistics. 2nd rev. Ed., Oxford University Press
Utsu T (1999) Representation and analysis of the earthquake size distribution: A historical review and some new approaches. Pure Appl. Geophys. 155: 509-535
Youngs RR, Abrahamson N, Makdisi FI, Sadigh K (1995) Magnitude-Dependent Variance of Peak Ground Acceleration. Bull. Seism. Soc. Am.  85: 4, pp. 1161-1176.

# Appendix

## *A1 - An inappropriate approach to model selection*

Scherbaum et al. (2004) formulated the criterion for model selection, which is the median of the statistic *LH*, defined with (symbols according to the reference)

$$LH(Z_0) = 2\left[1 - \Phi\left(|Z_0 / \sigma_0|\right)\right] \tag{A1}$$

wherein $Z_0$ is the estimated residual (here $\xi$) and its estimated standard deviation $\sigma_0$. $\Phi$ is the CDF of the standard normal distribution; a normal distributed $Z_0$ is desired resp. assumed. The smaller the value |*Median(LH)*-0.5|, the better is the model. The problem is that |*Median(LH)*-0.5|=0 for different distributions of $Z_0$. Examples are shown in Fig.A1. The criterion does not work.

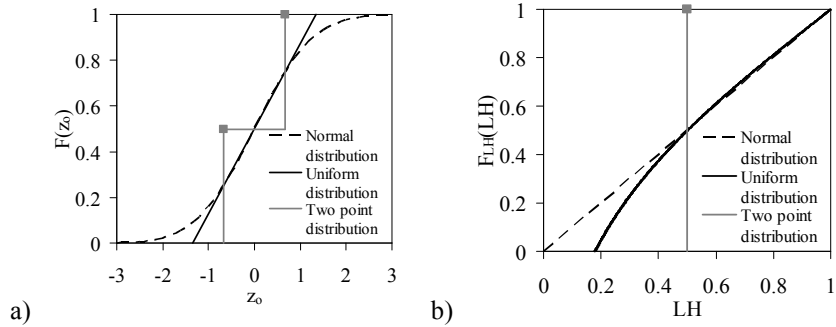

a)                                    b)

Fig.A1: Measure *LH* of Scherbaum et al. (2004) for $Z_0$ with different distributions: a) CDFs of $Z_0$ b) resulting CDFs of *LH* according to Eq.(A8), the median of *LH* is 0.5 in every case (uniform distribution: -1.3487≤ $Z_0$ ≤1.3487; two-point distribution: $z_{01}$=-0.6745 and $z_{02}$=0.6745)

## *A2 – Numerical research of distributions of combinations of horizontal components*

The horizontal components *(Y₁,Y₂)* of ground motion are combined for some GMRs as geometric mean, arithmetic mean or by a vectorial addition (s. Douglas 2003, Sec.6). The possibility of a logarithmic normal distribution of resulting $\varepsilon_0$ is researched here numerically. Therein, we assume Gumbel distributed components *($\varepsilon_{01},\varepsilon_{02}$)*. The other components of the GMR are not considered here because they scale *($\varepsilon_{01},\varepsilon_{02}$)* simultaneously. The dependence structure of *($\varepsilon_{01},\varepsilon_{02}$)* is assumed to be of a bivariate normal distribution (Mari and Kotz 2001, Sec. 4.4) and is quantified by the correlation coefficient *R*. We simulate pairs of horizontal component $\varepsilon_{01}$ and $\varepsilon_{02}$ with this copula and the Gumbel distributions as marginal and combine them. We Monte Carlo simulate a large sample (10000) of such combinations and check its logarithm for normal distribution with the Anderson-Darling test according to Stephens (1986) for a significance level α=5%. We repeat this procedure 100 times and get a share of rejected assumptions to be normal distributed. This share should be around 5%; otherwise, the considered combinations of Gumbel distributed maxima are not log-normally distributed. This is the case according to the results in Tab.A1.

Tab.A1: Share of rejections [%] for test of normality (estimated standard error in brackets; sample size n=10000, 100 repetitions, parameter b=1 of Gumbel distribution, s. Eq.(8b))

| Parameter *a* (variation coefficient) | 2.8 (0.380) | | | 3.3 (0.331) | | | 3.8 (0.293) | | |
|---|---|---|---|---|---|---|---|---|---|
| Correlation *R* | 0.2 | 0.5 | 0.8 | 0.2 | 0.5 | 0.8 | 0.2 | 0.5 | 0.8 |
| Geometric mean | 84 (3.7) | 93 (2.6) | 86 (3.5) | 55 (5) | 70 (4.6) | 76 (4.3) | 100 | 100 | 100 |
| Vectorial addition | 23 (4.2) | 57 (5) | 84 (3.7) | 95 (2.2) | 87 (3.4) | 77 (4.2) | 100 | 100 | 100 |
| Arithmetic mean | 42 (4.9) | 61 (4.9) | 80 (4) | 67 (4.7) | 77 (4.2) | 82 (3.8) | 100 | 100 | 100 |

## *A3 – Details of the constructed situation of seismicity*

The constructed source region and the considered site **s** are depicted in Fig.A2. The truncated exponential distribution for the magnitudes is formulated according to Cosentino et al. (1977) with

$$F_m(m) = (1 - \exp(-\beta_m(m - m_{\min}))) / (1 - \exp(-\beta_m(m_{\max} - m_{\min}))), \quad m_{\min} \le M \le m_{\max}, \qquad (A2)$$

wherein $\beta_m$ is a scale parameter, $m_{max}$ is the upper bound magnitude and $m_{min}$ is the smallest considered magnitude. We set $m_{min}$=4 and $\beta_m$=2.3 (s. Utsu 1999). The maximum magnitude $m_{max}$ depends on the investigated variant. The annual seismicity is set to $\nu$=4.4/600²[km⁻²], which means that 4.4/600² earthquakes with $M{\ge}4$ occur per km² in the source region (Fig.A2).

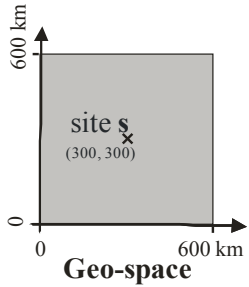

Fig.A2: Constructed source region with uniform seismicity and considered site in the geo-space.

## *A4 – Details of the simulations of Sec.5.4*

We assume the following for the Monte Carlos Simulation of sample in Sec.5.4. The beta distribution is applied to simulate a sample of random magnitude $m$ which is generally written with (s. Johnson et al. 1995)

$$f(x) = (x/(b-a))^{p-1} (1 - x/(b-a))^{q-1} \Gamma(p)\Gamma(q) / ((b-a)\Gamma(p+q)), a \le x \le b, \; p > 0, q > 0 . (A3)$$

The parameters for the beta distributed magnitude $m$ are listed for all variants in Tab.A2. The real epicenter distance is also simulated by a beta distribution with $b$=0 and with parameter $a$

$$a = cM^d . \qquad (A4)$$

The parameters $c,d, p$ and $q$ of the variants are listed in Tab.A3.

Tab.A2: Parameters for the constructed beta distribution of real magnitudes $M$

| Parameter | Variant of Tab.2 | | | |
|---|---|---|---|---|
| | **#1** | **#2** | **#3** | **#4** |
| *a* | 3.5 | 3 | 4 | 3 |
| *b* | 7 | 7.5 | 7.5 | 7.5 |
| *p* | 2 | 2 | 1 | 1 |
| *q* | 2 | 2 | 1 | 1 |

Tab.A3: Parameters for the constructed beta distribution of real epicenter distance $D$

| Parameter | Variant of Tab.2 | | | |
|---|---|---|---|---|
| | *#1* | *#2* | *#3* | *#4* |
| *p* | 2 | 2 | 1.7 | 1.7 |
| *q* | 2 | 1 | 1 | 1 |
| *c* | 200 | 100 | 200 | 20 |
| *d* | 0 | 0.5 | 0 | 1 |

### A5 – Details of the modeling and estimations of Sec.6

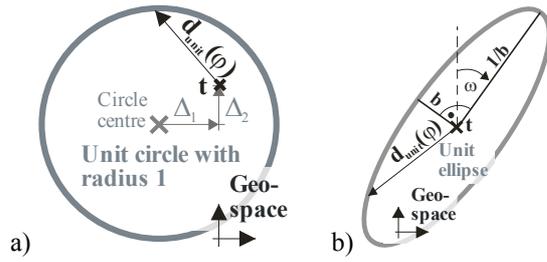

a)  b)

Fig.A3: Unit-isolines for the point source models of Sec.6.2: a) eccentric circle, b) ellipse with azimuth ω

Tab.A4: Estimated Parameters of the best variant (smallest AIC) of the different approaches and data sets according to Sec.6.2 (E and W geo-coordinates of **t**; $\Delta_X$, $\Delta_y$, $b$ and $\omega$ according to Fig.A3; $\theta_i$ according to Eq.(5))

| Approach | | Distance (to) | Para-meter | Earthquake # | | | | | | | | |
|---|---|---|---|---|---|---|---|---|---|---|---|---|
| | | | | 127 | 113 | 137 | 118 | 30 | 163 | 126 | 161 | 136 |
| Point source | Iso. | Seismol. Epicenter | $\theta_2$ | 0.00 | 0.000 | 0.004 | 0.00 | 0.00 | 0.00 | 0.00 | 0.00 | 0.01 |
| | | | $\theta_3$ | 1.16 | 1.00 | 0.79 | 1.00 | 1.70 | 1.62 | 1.00 | 0.54 | -0.08 |
| | | | $h$ | 9.49 | 10.79 | 6.76 | 17.48 | 13.00 | 15.20 | 34.21 | 9.10 | 15.00 |
| | | | $\theta_0$ | 2.43 | 1.19 | 1.30 | 2.17 | 4.18 | 2.79 | 1.63 | -2.34 | -1.60 |
| | | Estim. Epicenter | $\theta_2$ | 0.00 | 0.00 | 0.012 | 0.00 | 0.00 | 0.00 | 0.00 | 0.00 | 0.01 |
| | | | $\theta_3$ | 1.15 | 1.00 | 0.25 | 1.00 | 1.79 | 1.76 | 1.00 | 1.00 | -0.27 |
| | | | $h$ | 8.72 | 8.80 | 6.76 | 17.48 | 13.00 | 15.20 | 34.21 | 9.10 | 15.00 |
| | | | $\theta_0$ | 2.40 | 1.13 | -0.25 | 2.17 | 4.59 | 3.45 | 1.64 | -0.28 | -2.19 |
| | | | W | -118.55 | -118.12 | 120.90 | -121.95 | -118.35 | -116.52 | -116.72 | -116.70 | 30.09 |
| | | | N | 34.25 | 34.03 | 24.11 | 37.05 | 34.36 | 33.45 | 34.18 | 34.35 | 41.22 |
| | An-iso. | Estim. Epicenter | $\theta_2$ | 0.00 | 0.00 | 0.00 | 0.00 | 0.00 | 0.00 | 0.00 | 0.00 | 0.005 |
| | | | $\theta_3$ | 1.00 | 1.00 | 2.38 | 1.00 | 1.75 | 1.47 | 0.66 | 1.00 | 0.59 |
| | | | $h$ | 6.95 | 14.60 | 36.25 | 17.48 | 13.00 | 15.20 | 0.10 | 9.10 | 15.00 |
| | | | $\theta_0$ | 1.87 | 1.36 | 8.21 | 2.16 | 4.43 | 2.31 | 0.04 | -0.11 | 1.27 |
| | | | W | -118.49 | -118.20 | 120.98 | -121.88 | -118.41 | -116.35 | -116.92 | -116.95 | 30.86 |
| | | | N | 34.37 | 33.88 | 24.06 | 37.04 | 34.44 | 33.66 | 34.23 | 34.29 | 40.84 |
| | | | $\Delta_X$ | 0.22 | -0.38 | - | 0.13 | -0.13 | 0.35 | -0.34 | -0.43 | 0.34 |
| | | | $\Delta_y$ | 0.44 | -0.75 | - | -0.05 | 0.20 | 0.52 | 0.17 | -0.24 | -0.39 |
| | | | $b$ | - | - | 1.10 | - | - | - | - | - | - |
| | | | $\omega$ | - | - | 0.76 | - | - | - | - | - | - |
| Constructed distance | | Joyner-Boore | $\theta_2$ | 0.000 | 0.008 | 0.000 | 0.000 | 0.000 | - | - | - | 0.010 |
| | | | $\theta_3$ | 0.852 | 1.00 | 0.34 | 0.61 | 1.41 | - | - | - | 0.104 |
| | | | $h$ | 5.73 | 14.60 | 0.95 | 3.03 | 5.79 | - | - | - | 15.000 |
| | | | $\theta_0$ | 0.997 | 1.15 | -0.79 | 0.28 | 2.60 | - | - | - | -1.192 |
| | | Campbell | $\theta_2$ | 0.000 | 0.004 | 0.000 | 0.000 | 0.000 | - | - | - | 0.010 |
| | | | $\theta_3$ | 1.000 | 1.00 | 0.67 | 0.65 | 1.52 | - | - | - | 0.109 |
| | | | $h$ | 5.621 | 0.10 | 0.10 | 0.10 | 6.44 | - | - | - | 15.000 |
| | | | $\theta_0$ | 1.648 | 1.17 | 0.34 | 0.44 | 3.17 | - | - | - | -1.155 |
| | | RmsD | $\theta_2$ | 0.000 | 0.008 | 0.000 | 0.000 | 0.000 | - | - | - | 0.010 |
| | | | $\theta_3$ | 1.233 | 1.39 | 0.63 | 1.00 | 1.79 | - | - | - | -0.016 |
| | | | $h$ | 0.100 | 14.60 | 6.76 | 0.10 | 13.00 | - | - | - | 15.000 |
| | | | $\theta_0$ | 2.772 | 2.81 | 0.96 | 2.09 | 4.58 | - | - | - | -1.337 |
| | | ClstD | $\theta_2$ | 0.000 | 0.000 | 0.000 | 0.000 | 0.000 | - | - | - | 0.010 |
| | | | $\theta_3$ | 1.000 | 1.00 | 0.41 | 0.65 | 1.51 | - | - | - | 0.108 |
| | | | $h$ | 5.621 | 0.10 | 1.59 | 0.10 | 6.65 | - | - | - | 15.000 |
| | | | $\theta_0$ | 1.648 | 1.17 | -0.52 | 0.44 | 3.08 | - | - | - | -1.158 |